\RequirePackage[hyphens]{url}
\documentclass[reprint,aps,prper,twocolumn]{revtex4-2} 
\usepackage{graphicx}
\usepackage{hyperref}
\usepackage{footnote}
\usepackage{amsmath,amssymb}
\usepackage{multirow}
\usepackage{tabularx}
\usepackage[table]{xcolor}
\usepackage{booktabs} 
\usepackage{cleveref}
\usepackage{makecell}
\usepackage{supertabular}
\usepackage{comment}
\usepackage{enumitem} 
\usepackage{lineno}
\newcommand{\code}[1]{\texttt{#1}}
\usepackage{float}
\usepackage{array}
\usepackage{pdfpages}
\newcolumntype{P}[1]{>{\centering\arraybackslash}p{#1}}
\newcolumntype{M}[1]{>{\centering\arraybackslash}m{#1}}

\makeatletter
\AtBeginDocument{\let\LS@rot\@undefined}
\makeatother

\begin{document}

\title{Affordances and Challenges of Incorporating a Remote, Cloud-accessible Quantum Experiment into Undergraduate Courses}
\date{\today}

\author{Victoria Borish}
\email[]{victoria.borish@colorado.edu}
\author{H. J. Lewandowski}

\affiliation{Department of Physics, University of Colorado, Boulder, Colorado 80309, USA}
\affiliation{JILA, National Institute of Standards and Technology and University of Colorado, Boulder, Colorado 80309, USA}

\begin{abstract}
As quantum technologies transition from the research laboratory into commercial development, the opportunities for students to begin their careers in this new quantum industry are increasing. With these new career pathways, more and more people are considering the best ways to educate students about quantum concepts and relevant skills. In particular, the quantum industry is looking for new employees with experimental skills, but the instructional labs, capstone projects, research experiences, and internships that provide experiences where students can learn these skills are often resource-intensive and not available at all institutions. The quantum company, Infleqtion, recently made its online quantum matter machine Oqtant publicly available, so people around the world could send commands to create and manipulate Bose-Einstein condensates and receive back real experimental data. Making a complex quantum experiment accessible to anyone has the potential to extend the opportunity to work with quantum experiments to students at less-resourced institutions. As a first step in understanding the potential benefits of using such a platform in educational settings, we collected data from instructors and students who were interested in using, or had used, Oqtant. In this study, we investigate instructors' views about reasons they would like to use Oqtant and challenges they would face in doing so. We also provide a concrete example of how Oqtant was used in an upper-division undergraduate quantum mechanics course and the instructor's perception of its benefits. We complement this with the student perspective, discussing student experiences interacting with Oqtant in their course or through think-aloud interviews outside of a course. This allows us to investigate the reasons students perceive Oqtant to be a real experiment even though they never physically interact with it, how Oqtant compares to their other experimental experiences, and what they enjoy about working with it. These results will help the community consider the potential value for students of creating more opportunities to access remote quantum experiments.
\end{abstract}

\maketitle

\section{Introduction}\label{sec:intro}

Due to the increasing prevalence of quantum technologies, the United States and other countries are putting many resources into educating students about quantum information science and engineering (QISE) \cite{nationalQuantumInitiativeAct,NQIsupplement,riedel2019europe}. Educational efforts at different levels are already underway, not only to add more modern topics to the physics curriculum, but also to better prepare students for, and broaden access to, quantum careers \cite{perron2021quantum, meyer2022todays, kaur2022defining, hasanovic2022quantum, stadermann2019analysis, seskir2022quantum,asfaw2022building,aiello2021achieving}. This education may include both conceptual learning of key topics in addition to skill development, which will help students learn how to work with the different experimental platforms used in quantum technologies \cite{fox2020preparing, greinert2024advancing, hughes2022assessing, greinert2023future}.

Although experimental skills are an important part of all physics students' undergraduate education \cite{kozminski2014aapt,joint2016phys21} and are valued by employers in the quantum industry \cite{fox2020preparing, aiello2021achieving, asfaw2022building}, opportunities to learn these skills are not available to all students. Quantum companies are looking for students with technical skills, including both specific hands-on skills such as aligning lasers and skills that span various experimental platforms such as coding, data analysis, and troubleshooting \cite{fox2020preparing, aiello2021achieving}. There are various ways for undergraduate students to learn experimental skills including lab courses, capstone projects, research experiences (traditional or course-based \cite{auchincloss2014assessment, buchanan2022current, merritt2024physics}), and internships. These experiences also provide opportunities for students to engage in authentic scientific practices \cite{smith2020direct, dutson1997review,holmes2016examining,oliver2023student}. However, all of these options are resource intensive. There are often more interested students than intern or research positions, and even lab courses require significant funds and instructor expertise to be able to include quantum experiments that are at the forefront of modern technologies \cite{asfaw2022building}. As part of the many calls to reduce barriers to participation in QISE \cite{NQIsupplement,meyer2024disparities,rosenberg2024science,bennett2024investigating}, there is a need to increase student access to quantum experiments.

One option to teach relevant experimental skills to a wide set of students is to use remote, cutting-edge quantum experiments. In the fall of 2023, the quantum company, Infleqtion, made publicly available their online Bose-Einstein condensate (BEC) \cite{anderson1995observation} experiment, which anyone with an internet connection can use to create and manipulate quantum matter \cite{oqtantWebsite}. This experimental platform provides users the opportunity to control the experimental apparatus and receive and analyze data in a manner similar to that which is done in modern research labs \cite{tingle2024oqtant}. Oqtant, therefore, has the possibility to provide students the opportunity to learn experimental skills relevant for the quantum industry such as controlling an ultracold atom experiment or measuring and analyzing quantum states. There have been recent efforts to make physics experiments accessible to anyone with an internet connection \cite{castano2024deploying, remoteGlowDischargeExpt, IBMquantumComputing}, and Oqtant is one of the first examples of a publicly-available quantum experiment that does not abstract away all elements of the hardware. Although the future of Oqtant as a commercially sustainable product is uncertain, it is important to investigate educational experiences with Oqtant as an example of the utility of remote quantum experiments to educate physics students and the quantum workforce.

Since Oqtant provides a new type of educational opportunity, we investigate broadly the affordances and challenges that come with using the platform in an undergraduate course, considering the perspectives of both instructors and students. We surveyed instructors interested in using Oqtant in their courses and interviewed one instructor who had incorporated Oqtant into their course to answer the following research questions:
\begin{enumerate}[label=I\arabic*.,ref=I\arabic*]
    \itemsep0em
    \item \label{RQ:learningGoals} What learning goals would instructors have for incorporating Oqtant into their courses?
    \item \label{RQ:challenges} What challenges would instructors face when implementing Oqtant in their courses?
    \item \label{RQ:support} What kinds of support would instructors need to implement Oqtant in their courses?
\end{enumerate}
We additionally detail one implementation of Oqtant in an upper-division quantum mechanics course as a concrete example, answering the research questions:
\begin{enumerate}[label=I\arabic*.,ref=I\arabic*]
    \setcounter{enumi}{3}
    \itemsep0em
    \item \label{RQ:implementation} How could Oqtant be implemented in an undergraduate quantum mechanics course?
    \item \label{RQ:outcomes} What were the instructor's perceived outcomes of using Oqtant in a quantum mechanics course?
\end{enumerate}

To obtain the student perspective, we performed think-aloud interviews outside of a course setting and additionally studied course materials from students in the quantum mechanics course described in questions \labelcref{RQ:implementation,RQ:outcomes}. These data sources allow us to answer the following research questions:
\begin{enumerate}[label=S\arabic*.,ref=S\arabic*]
    \itemsep0em
    \item \label{RQ:real} Do students feel like they are working with a real experiment while working with Oqtant?
    \item \label{RQ:comparison} How does Oqtant compare with other experimental experiences students have had?
    \item \label{RQ:like} What do students enjoy and not enjoy about working with Oqtant?
\end{enumerate}
This study is not an evaluation of the students or the course; rather, the combined data from all of these questions allow us to understand the potential opportunities and difficulties of using Oqtant for educational purposes. We hope this example of our developed educational materials and the way they were implemented in an upper-division course along with the consequent research results will help the community consider the implications of using remote quantum experiments to make quantum education more accessible and guide future research into the efficacy of this approach.

The rest of this paper is organized as follows. In Sec.~\ref{sec:bg}, we provide additional background information about the demand for students with experimental quantum skills, low-resource alternatives to quantum experiments, and the experimental platform Oqtant. We then detail our methodology including the various data sources, data analysis techniques, and limitations of this study in Sec.~\ref{sec:methods}. Next, we present the results in Sec.~\ref{sec:results}, broken into instructor perceptions of using Oqtant, one concrete implementation, and student perceptions. Finally, we conclude in Sec.~\ref{sec:conclusions}, summarizing our results and discussing how to sustainably create opportunities for remote quantum experiments in educational settings.

\section{Background}\label{sec:bg}

Although there has been education research on student learning of quantum mechanics for decades, only recently has there been a focus on student learning of QISE \cite{kohnle2017interactive, meyer2022todays,hu2024investigating} and the needs of the quantum workforce \cite{fox2020preparing, hughes2022assessing, greinert2023future, greinert2024advancing, kaur2022defining}. Most of the research on student learning focuses on students' conceptual understanding even though there is a recognized need to provide students experiences with quantum experiments \cite{asfaw2022building,borish2023seeing}. In this section we discuss the importance of quantum experimental skills to the quantum industry and the hands-on quantum experiences that are already being incorporated into undergraduate courses. We then describe less resource-intensive options to help students learn about quantum experiments, including simulations, virtual labs, and remote experiments. Lastly, we describe the capabilities of Oqtant and the ways students can interact with it.

\subsection{Quantum experimental skills}

The utility of students interacting with quantum experiments has shown up both in discussions of students interested in pursuing a career in the quantum industry \cite{fox2020preparing, aiello2021achieving, hasanovic2022quantum,asfaw2022building,greinert2024advancing} and within typical undergraduate courses \cite{lukishova2022fifteen, borish2022seeing,borish2023seeing, lukishova2024teach}. For example, many quantum companies value experimental skills more than a detailed understanding of quantum theory when looking for new hires \cite{fox2020preparing,aiello2021achieving}. Companies cite a wide variety of beneficial experimental skills, including ones that require students to work hands-on with an apparatus, such as aligning laser systems or working with electronics \cite{fox2020preparing, asfaw2022building, aiello2021achieving, greinert2024advancing}. Other relevant skills, such as programming to control the experimental apparatus, collecting and analyzing data, and documenting and reporting \cite{fox2020preparing, aiello2021achieving}, may not require hands-on interaction with the apparatus. Since quantum technologies involve a wide variety of experimental platforms, it can also be useful to provide students opportunities to practice with different types of platforms, including ultracold neutral atoms \cite{asfaw2022building}.

Students can obtain hands-on experiences with quantum experiments through lab courses, capstone projects, research experiences, or internships. Various quantum experiments are already incorporated into many traditional quantum mechanics or upper-division lab courses \cite{pina2024united}. One example is a popular set of quantum optics experiments \cite{galvez2005interference, beck2012quantum, lukishova2022fifteen} that is often used to teach students various lab skills including optical alignment and data acquisition and analysis \cite{borish2023implementation}. These experiments have been shown to produce other positive student outcomes such as confirming students' beliefs that quantum mechanics describes the physical world \cite{borish2023seeing}, improving students' conceptual understanding, and motivating students to pursue a career in quantum optics or quantum information \cite{lukishova2022fifteen}. Quantum industry senior capstone courses are being created \cite{oliver2024education}, but research on skills students develop in them is ongoing. We are not aware of other research specifically studying the extent to which students have gained experimental skills from quantum research experiences or internships, although they would likely be similar to those of undergraduate research experiences or internships in general \cite{hunter2007becoming}.

Although these hands-on experiences provide benefits to students, there are also various barriers to their implementation, leading to them not being available to all students. These barriers include the need for often expensive equipment, the necessary infrastructure to support the experiments, sufficient time for instructors or mentors to prepare for and support their students, and instructor expertise during development and maintenance of the experiment \cite{aiello2021achieving, asfaw2022building}. These challenges can also make it difficult to scale the use of experiments to large class sizes \cite{asfaw2022building, aiello2021achieving}. Some studies have proposed remote experiments as an alternative to in-person experiments when they are not available \cite{hasanovic2022quantum, asfaw2022building}, including experiments with BECs that require substantial local expertise \cite{asfaw2022building,tingle2024oqtant}. The significant resources needed to provide students exposure to quantum experiments, along with the importance of the skills students could gain from working with them, makes it important to investigate methods of teaching these skills with fewer resources.

\subsection{Less resource-intensive alternatives to quantum experiments} \label{sec:bgAlternatives}

There have been many interactive simulations created to teach students concepts related to quantum experiments without the need for experimental apparatus \cite{mckagan2008developing,ahmed2022student,kohnle2015enhancing, malgieri2014teaching, marshman2022quilts}. There is existing evidence that engaging with simulations can improve students' conceptual understanding of quantum topics \cite{malgieri2014teaching,kohnle2015enhancing,marshman2022quilts,marshman2016interactive}. While many of these simulations focus on one specific type of experimental context (e.g., a Stern-Gerlach apparatus or a Mach-Zehnder interferometer), there are other types called ``virtual labs'' that simulate results of experimental setups students design in the simulation. For example, students can begin with the virtual equivalent of an empty optical breadboard and place various optical elements (such as lasers, waveplates, beamsplitters, and detectors) to design custom quantum optics experiments \cite{la2021virtual,migdal2022visualizing}. These simulations allow students to visualize aspects of physics that are impossible to visualize in actual experiments and are easily scalable for large class sizes \cite{migdal2022visualizing}. However, simulations do not provide students opportunities to gain hands-on skills such as optical alignment, although students could learn the roles of each piece of equipment, how to decide which pieces of equipment are needed in a given experiment, and data analysis techniques \cite{seskir2022quantum}. We are not aware of any studies specifically investigating experimental skills students gain with these simulations, as there are no standard methods of assessing many types of experimental skills.

Another less-resource-intensive method of including ideas from quantum experiments into courses is to incorporate photos and videos of an actual apparatus and real experimental data into student activities. This requires educators to have access to one version of the apparatus in order to develop the materials, but the physical apparatus is not needed after that so activities can easily be scaled for many students. Video clips of a collection of real data combined with simulations have been shown to improve students' quantum reasoning about concepts such as interference \cite{waitzmann2024testing}. Interactive screen versions of quantum optics experiments \cite{bronner2009interactive}, which allow students some choice in experimental parameters, although they still see pre-recorded data, have been shown to help students avoid using deterministic reasoning about quantum objects \cite{bitzenbauer2021effect}. These options allow students to obtain real experimental data that they have some control over while decreasing the required resources to only a single apparatus that could exist at a different institution. These activities have been used primarily to improve students' conceptual learning \cite{bitzenbauer2021effect, waitzmann2024testing}; however, similar activities may be able to improve students' data analysis skills and views about how knowledge comes from experimental observations. 

Some institutions have also converted their previously in-person quantum experiments to be able to be accessed remotely \cite{galvez2021remote, UCSBremoteLabs}. These experiments allow students to remotely control various components of the instructional lab hardware via the internet, and the students can see the effects on their actions on the apparatus via live feeds from cameras in the lab. Compared with interactive screen experiments, remote experiments provide students more options for how they interact with the experiment (as the settings they choose need not be already measured by an educator), but they do not allow for a large number of students to work with them concurrently. Compared with in-person experiments, remote experiments more easily allow multiple groups to work with the same experiment sequentially, may decrease student concerns about breaking equipment, and increase access for students who are not able to physically be in the lab at a given time. Remote labs may also provide students the opportunity to learn how to design experimental procedures and troubleshoot experiments remotely \cite{galvez2021remote}.

Quantum industry also provides access to some remote quantum experiments via publicly-accessible cloud quantum computers (e.g., Refs.~\cite{IBMquantumComputing, IonQquantumCloud, QuEra}). These have begun appearing in educational programs, including informal education for high school students \cite{qubitByQubit, tappert2019experience}, an online global summer school \cite{QiskitSummerSchool, singh2021preparing}, and undergraduate and graduate courses \cite{tappert2019experience, brody2023testing}. The various quantum computers have different price structures, interfaces, and provided tutorials and educational materials, varying the ease of use for students. In addition to teaching students about quantum computing topics, cloud quantum computers can be used to teach about a variety of other quantum topics, such as testing Bell's Inequality \cite{brody2023testing} and modeling quantum sensors \cite{tran2022modeling}. Although educational experiences with quantum computers provide students experience with authentic quantum devices, many of the educational materials created to use with them focus more on quantum concepts than experimental skills. 

\subsection{The cloud-accessible quantum matter experiment, Oqant}

Oqtant is the only publicly-available quantum experiment hosted by a quantum company that is not a quantum computer of which we are aware. Unlike quantum computers, where users may think entirely in terms of quantum circuits without knowledge of the underlying hardware, Oqtant emphasizes the system under study, a BEC, and allows users to set values for the controls available with the hardware. A BEC is a state of matter in which the constituent atoms of a gas have been cooled until a large fraction of them enter the lowest-energy state. 
At that point, the interparticle spacing is comparable to the de Boglie wavelength, so the atoms act as a single macroscopic quantum object and exhibit quantum properties such as interference \cite{pitaevskii2003bose}. Oqtant therefore provides users a unique opportunity to engage with, and visualize, fundamental physics while working with an experimental platform similar to some types of modern quantum technologies \cite{tingle2024oqtant}.

\begin{figure*}
    \centering
    \includegraphics[width=\linewidth]{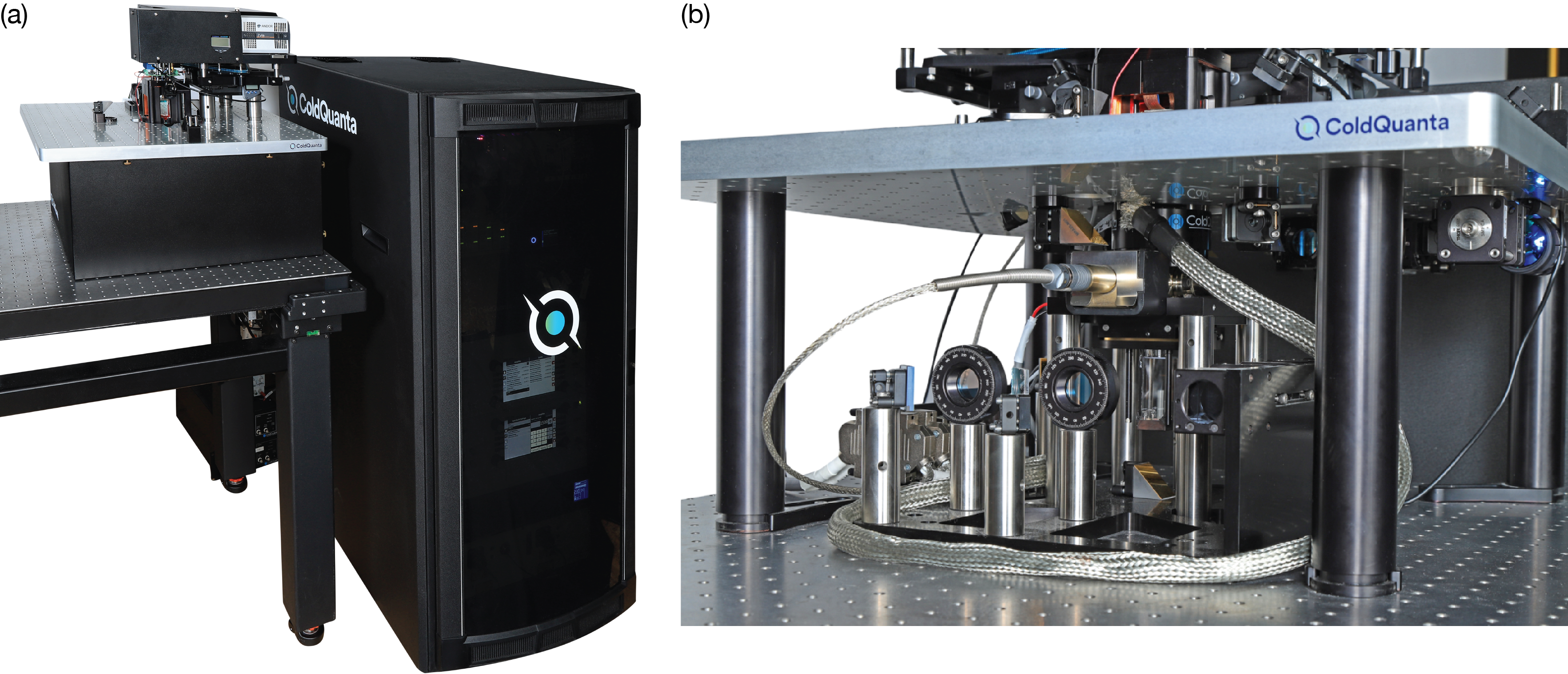}
    \caption{Photos of (a) the entire Oqtant apparatus and (b) a close-up of the optics surrounding the vacuum chamber in which the atoms are located. Photos taken by the Oqtant team.}
    \label{fig:hardware}
\end{figure*}

Oqtant can be accessed from almost anywhere via the internet, while the hardware resides in the quantum company Infleqtion's office in Colorado. The apparatus itself has similar features to that of many other BEC experiments in research labs around the world \cite{salim2013high, lewandowski2003simplified, inguscio1999bose}, but it has been optimized for the required stability, low maintenance, and automation needed for it to be available to the public. A photo of the apparatus can be seen in Fig.~\ref{fig:hardware}, where the optics are compact and all of the electronics and lasers are placed in racks. The purpose of Oqtant is to provide researchers and educators a platform they can use to access ultracold atoms without the money and time it takes to build such an experiment themselves \cite{tingle2024oqtant,oqtantWebsite}.

In an experimental run with Oqtant, rubidium atoms in an ultra-high vacuum chamber undergo various cooling and trapping stages, with the users controlling the final stages. At the start of each run, the atoms are cooled and trapped with lasers and a magnetic field. The final cooling stage, evaporative cooling in a magnetic trap, is where users are first allowed to control the experimental parameters. Users have the option to adjust the frequency, power, and time intervals of a sequence of applied radio frequency radiation that cools the atoms to the point where some begin to form a BEC. Users can either image the BEC immediately or wait until after applying a spatially-localized far-detuned laser, which repels atoms from the regions with that optical field. This can be used, for example, to split the BEC into different sub-ensembles of BECs. Users can decide to image the atoms while the atoms remain in the magnetic trap or they can release the atoms from the trap and image them after the atoms have expanded for a variable amount of time \cite{tingle2024oqtant,oqtantWebsite}. The images of the atoms provide the atomic density profile, which allows for calculations of the temperature of the atoms, the total number of atoms, and the number of atoms that have formed a BEC. These features allow users to investigate topics related to fundamental quantum phenomena including superposition, interference, and tunneling \cite{tingle2024oqtant,pitaevskii2003bose}.

There are two ways users can access Oqtant, either with a web application or a Python application programming interface (API). The web application is easier to use, and the Python API allows for additional ways to control the experiment. Through either method, users submit ``jobs'' that contain the experimental parameters they want to run on the experiment. These jobs enter a queue and run when the experiment is online. Once a job has completed, the user can access it to obtain the resulting image of the atoms (as a visual image or a matrix) along with the basic data analysis provided by Oqtant (the calculated atom number and temperature). Infleqtion provides sample Jupyter notebooks containing tutorials and demonstrations to help new users and offers various methods of support (e.g., by email or Slack) if users run across issues \cite{tingle2024oqtant,oqtantWebsite}.

Anyone can obtain a free account by signing up on Oqtant's website, but jobs can only be run when the experiment is online. Since Oqtant was first online in fall 2023, users with free accounts have had access to between 10--20 jobs per day (with a maximum of 100 jobs per month). Users who wanted additional jobs could purchase more, which increases their daily limit and designates them as priority jobs that jump to the front of the queue. In its first year, Oqtant was online from 10 am to 3 pm Mountain time on Tuesdays, Wednesdays, and Thursdays every other week. In September 2024, Oqtant announced it will indefinitely pause its services beginning November 2024, and it is unknown if and when it will be available again in the future \cite{oqtantWebsite}.

\section{Methodology}\label{sec:methods}

In order to study a cloud-accessible quantum experiment, which is novel both in terms of the technology itself and the opportunities for education and education research, we collected and analyzed several different types of data from various perspectives. This allows us to provide a broad overview of the possibilities of using Oqtant in an educational setting, so future research can narrow in on more specific details of implementations. We begin this section detailing the data sources we used, which come from instructors interested in using Oqtant in their courses, as well as an instructor and students who worked with Oqtant using educational materials we developed. We then discuss how these data were analyzed with a thematic analysis, looking for the existence of themes. Finally, we discuss the limitations of this study and how those contributed to our analysis emphasis of investigating educational possibilities.

\subsection{Data collection}

We used several different data sources to answer our research questions, which are summarized in Table~\ref{tab:data}. In order to study student experiences with an experiment like Oqtant in a course, we needed to develop educational activities that could be used by the students. To determine the structure of activities that would fit into existing courses, we surveyed instructors about their interest in using Oqtant in their courses. The data from this survey were used both to develop the student activities and to answer some of our research questions. Once the activities were developed, we performed think-aloud interviews with students using these activities both to improve the activities and to understand student experiences while working with Oqtant. We then implemented the finalized activities in an upper-division quantum mechanic course. Reflection questions on the students' completed activities along with an interview of the instructor provided us information about the use of Oqtant within a course setting.

\begin{table*}[htbp]
\centering
\caption{Data sources used in this study with the number of participants and their institution type for each part along with the research questions (RQs) each were used to answer. The students were enrolled either at research-intensive R1 institutions or a primarily-undergraduate institution (PUI).}
\begin{tabular}{l l l} 
    \hline \hline
    \textbf{Data source} & \textbf{Participants and institutions} & \textbf{RQs} \\
    \hline 
    Instructor survey & 29 instructors from 28 institutions & \labelcref{RQ:learningGoals,RQ:challenges,RQ:support}\\
    Instructor interview & 1 instructor at a Hispanic-serving PUI & \labelcref{RQ:learningGoals,RQ:challenges,RQ:support,RQ:implementation,RQ:outcomes}\\
    Sets of students' completed activities & 5 students at R1 institutions, 7 students at a PUI &
    \labelcref{RQ:real}--\labelcref{RQ:comparison} \\
    Student think-aloud interviews & 5 students at R1 institutions & \labelcref{RQ:real,RQ:comparison,RQ:like} \\
    \hline \hline
\end{tabular}
\label{tab:data}
\end{table*}

\subsubsection{Instructor survey}

We recruited instructors for the instructor survey through a variety of methods, including via email and conversations at conferences. We posted a link to the survey in the newsletter of the Advanced Physics Lab Association (ALPhA) \cite{alphaWebsite}, asking for responses from instructors teaching upper-division quantum mechanics courses, beyond-first-year (BFY) lab courses, or any other courses in which Oqtant could be incorporated. We additionally emailed 164 instructors who we knew were interested in hands-on quantum optics experiments, many of whom were connected to ALPhA in some way. We also presented a poster about our upcoming project with Oqtant (including a QR code to the survey) at the Conference on Laboratory Instruction Beyond the First Year and the American Association of Physics Teachers Summer Meeting in the summer of 2023. 

In total, 29 instructors from 28 unique institutions completed the survey. Of these institutions, six are four-year colleges, eight are master's degree granting institutions, 13 are PhD granting institutions (including four that are international), and we do not have information about the remaining one. One instructor submitted two responses, one each for two separate courses they teach, and one instructor submitted a single response for multiple disparate courses. 

The survey contained both open- and closed-response questions about the ways in which instructors would consider implementing Oqtant in their courses. These questions covered topics such as the courses in which instructors would want to use Oqtant, the amount of time they would be willing to dedicate to Oqtant, the learning outcomes they hope their students would achieve while working with Oqtant, their students' current knowledge about BECs, the concerns they have about implementing Oqtant in their course, and the support they would need to do so. The instructors were not familiar with Oqtant prior to filling out the survey, so we included a brief description of Oqtant along with links to access the platform at the start of the survey. The exact wording of the questions asked, as well as our description of Oqtant, is provided in the Supplemental Material \cite{SM}. 

\subsubsection{Educational materials developed}

We used responses to the instructor survey to aid in the development of educational materials that could fit into a variety of undergraduate courses. We designed the activities to be incorporated into upper-division lab or quantum mechanics courses, assuming minimal student prior knowledge of BECs. We did assume that students would have a basic knowledge of programming in Python including generating basic plotting commands, using \code{for} loops, and slightly adjusting already-written code. We developed a sequence of two activities, each designed to take students two to three hours to complete, with preparation activities students could work through on their own for around half an hour beforehand.

We employed educational best practices in the development of these activities \cite{holmes2023strategies}. Each of the activities and preparation activities included explicit learning goals, so the students would know what they were expected to learn from the activities \cite{simon2009value}. To encourage students to actively engage with the materials, we included questions at various points throughout the preparation activities to which students needed to provide a written response. We additionally added some metacognitive reflection questions throughout the activities to help with student sensemaking, as well as at the end of each activity to  encourage the students to think more broadly about their overall experience working with Oqtant. Although these activities needed to be somewhat structured due to the complexity of Oqtant, we provided students opportunities to make some decisions, such as the requested temperature of the atoms and the number of runs they took. We created two structured activities with the plan that they could be followed by a student-designed open-ended project.

Development of the two activities was an iterative process, where we went back and forth between the desired learning outcomes and the capabilities of, and data produced by, Oqtant. Our initial learning goals were chosen based off of the results from the instructor survey and our knowledge of experimental atomic physics and BECs. Extensive time experimenting with Oqtant allowed us to refine the learning goals so they would be feasible in a short period of time with a limited number of experimental runs. Both of the activities were structured as Jupyter notebooks in which the students could submit runs to Oqtant, analyze the resulting data, and respond to the reflection questions. The associated preparation activities consisted of a few pages of background reading, around four related questions to which the students needed to submit short responses, and short Jupyter notebooks the students could run to ensure their Python connection was working. The final versions of our developed activities can be found in Ref.~\cite{LewGroupWebsite}.

The first activity guides students to learn about absorption imaging of ultracold atoms (the process used to produce the images returned to the user by Oqtant) and understand how to use the images to obtain properties of the atoms. Students begin by analyzing already-taken data from Oqtant, using 2D Gaussian fits of the imaged atoms to obtain the number of imaged atoms and the temperature of the atomic cloud. Then, students have the opportunity to submit runs to Oqtant themselves, choosing a set of requested temperatures for the atoms. At the end of this activity, students may submit jobs that create BECs, but the focus is on understanding how to detect ultracold atoms and determine their properties. 

In the second activity, students focus on the creation, and some of the quantum aspects, of BECs. They begin by comparing the Gaussian fits they had been using in the first activity with bimodal fits, which better fit the atoms when a BEC is present, allowing them to consider different regimes and decide when one model works better then the other. Using Oqtant's built-in functionality, students then find the critical temperature, the temperature at which a BEC starts to form. They additionally investigate the parameters sent to the hardware for the final cooling stage, thereby making a connection between the frequency of the applied radio frequency radiation and the final temperature of the atoms. Finally, the students compare the aspect ratio of a BEC versus a non-condensed cloud of atoms after different expansion times. 

\subsubsection{Think-aloud interviews}

Once the first versions of the educational activities with Oqtant were completed, we began testing them with students using think-aloud interviews \cite{charters2003use}. These interviews occurred over Zoom while Oqtant was online, so the students could submit jobs and receive the results during the interview. All the students performed two think-aloud interviews, one for each of the activities, and additionally completed the respective preparation activities ahead of time. The think-aloud portions of the interviews were used to improve the activities, for example by re-wording any parts that were confusing. We also included some regular interview questions after the students had completed the activities, and students responses to those questions were used to answer our research questions about student experiences.

We attempted to recruit students from various institutions and ended up with five students from two institutions who completed the sequence of think-aloud interviews. We advertised this opportunity to the second semester quantum mechanics course at our own institution, as well as to 14 instructors we knew from the survey or through personal connections, prioritizing instructors at institutions that were less likely to have access to in-person atomic physics experiments. The students who signed up were enrolled at two different institutions: two were enrolled in a primarily white R1 institution and the other three were enrolled in a Hispanic-serving R1 institution. Of the five students, four identified as men and one as a woman. Two are white, one is Asian, and two are both white and Asian. All of the students were in their last or second to last year of their undergraduate studies, had taken at least one programming class, and had taken a quantum mechanics course or participated in a quantum research experience.

A week prior to each think-aloud interview, the students were sent the preparation activities. These were intended to last between 15--30 minutes, although some of the students spent longer on the first one because it included installation of Oqtant's Python package, which was a challenge for many students. Students recorded responses to the preparation activity questions and emailed them to us before the think-aloud interviews. 

During the think-aloud interviews, students spent approximately two hours working through the activities with Oqtant while screen-sharing their Jupyter notebook over Zoom. During this time, the interviewer prompted the students to explain their reasoning if they were not explaining what they were doing out loud and also answered the students' questions about the activity, as an instructor would. Many of the student questions related to Python code, so all of these questions were answered as students' coding ability was not the focus of our study and we wanted to ensure students could finish the activities in the allotted time. After the think-aloud portion, students were asked up to 15 minutes of additional questions related to potential learning outcomes and were also given the chance to suggest changes to the activities (see Ref.~\cite{SM} for the exact questions). Students were compensated with gift cards for their time.

The data we collected during these think-aloud interviews consist of students' responses to the preparation activity questions, students' completed Jupyter notebooks, and the transcripts of the interviews. From the notebooks, we analyzed only student responses to the broad reflection questions at the end. We also focused on the end portion of the think-aloud interviews, where the students discussed their responses to the concluding reflection questions and answered some additional questions after completing the activities. 

\subsubsection{Course implementation} \label{sec:DataCourse}

In parallel to performing the think-aloud interviews, we began discussing the implementation of these activities with an upper-division quantum mechanics course instructor. We partnered with an instructor with whom we had previously connected about quantum education and who had directly expressed interest in using our activities with Oqtant. This instructor wanted to include Oqtant in their upper-division quantum mechanics course at a Hispanic-serving primarily-undergraduate institution. We had various conversations with them early on in the Spring 2024 semester to ensure these activities would match with their intended learning goals. We provided the instructor the updated activities, as we had made some small changes to them after the think-aloud interviews. We additionally were in contact with the instructor as the activities were being implemented to answer a few additional questions the instructor had. The details of how these activities were incorporated into the course are discussed in Sec.~\ref{sec:implementation}.

Our data from the course consist of student course materials related to Oqtant from the majority of the students in the course, as well as an instructor interview. There were 12 students enrolled in the course, mostly seniors who were going to graduate at the end of that semester or the following one. Seven of the students gave us permission to use their course materials in this study, so we have seven sets of responses to the preparation activities and Jupyter notebooks. We attempted to interview the students at the end of the course to ask questions similar to what we did at the end of the think-aloud interviews, but none of the students signed up. We were able to interview the instructor, so our dataset also contains the transcript of an hour-long Zoom interview with the instructor that occurred soon after the end of the course.

\subsection{Data analysis}

We analyzed the student and instructor data separately. For the closed-response instructor survey questions, we divided the responses by course type (see Ref.~\cite{SM} for details) and present instructor views about a total of 31 courses. For the open-response survey questions, reflection questions, and interview data, we performed thematic coding analyses \cite{braun2006using, merriam2015qualitative} to identify themes across the participants, with discussions between the authors throughout the process to refine the themes and agree on classification of the quotes. For the instructor data, this consisted of identifying themes in the survey data, and then searching for those and other emergent themes in the instructor interview. These themes are presented in Tables~\ref{tab:challenges}--\ref{tab:support}. For the student data, we performed a thematic analysis of the interview transcripts and activity reflection questions from all 12 students. We chose to combine the student data from the think-aloud interviews and the quantum mechanics course because both sets of students worked through the same activities with Oqtant and our research questions do not depend on the specific context in which the activities were performed. Since our sample size is relatively small, we focus our analysis on the existence of themes instead of their prevalence. These themes are presented in Tables~\ref{tab:real}--\ref{tab:enjoy}.

To explain the students' and instructors' ideas, we chose the most representative or insightful quotes to present in this paper. There were misspelled words in some of the students' written responses, so we corrected those to minimize distraction. Because we do not have demographic information from all participants and to protect their anonymity, we use gender neutral pronouns for everyone.

\subsection{Limitations}

The primary limitation of this study is the small sample size of students. Although we tried to recruit more students for the think-aloud interviews, we could schedule the interviews only when Oqtant was online, which limited both the number of time slots we had available and the students who were free during those times. Some students may have had courses or other responsibilities that limited their ability to commit to at least two hours during the school week. The majority of the students in the quantum mechanics course provided permission to use their course materials, although we were not able to obtain data from the entire class. However, all of the claims in this paper are about the existence of ideas students may have, so although we are likely missing some ideas from students who were not able to participate in this study, the data here present valuable insights from a subset of students. Future work can investigate how these ideas may differ across disparate populations of students, as we do not have a large enough sample of students to make claims about such possible differences. The details of the course provided in Secs.~\ref{sec:DataCourse} and \ref{sec:implementation} can help identify how these results may generalize to other populations of students \cite{eisenhart2009generalization, robertson2018selection}.

Additionally, we studied student experiences with only one set of activities, which did require students to have at least some prior knowledge of programming in Python. We designed our research questions and focused our analysis on themes in the data that did not depend on the specific activities with which the students engaged; however, it is not possible to completely separate student experiences with Oqtant from the specific implementation. We have no reason to believe that some of the general themes we found in our data would be unique to the way the students in our study interacted with Oqtant, so we hope this study will motivate future work to better understand a wider range of student experiences with remote quantum experiments.

From the instructor side, all of the survey responses came from instructors who had not yet used Oqtant in their courses, whereas the interview came from an instructor who had already used Oqtant. It is important to understand the perspectives of both instructors who might use Oqtant, but have not yet (to understand why or why not instructors would consider incorporating a novel experiment like Oqtant into their courses) and instructors who have used Oqtant in their courses (to understand whether the perceived challenges and support needed went as expected or whether new challenges occurred). Due to there only being one instructor who had used Oqtant, we combined responses from the survey with the instructor interview when presenting the results in Sec.~\ref{sec:instructorPerceptions}.

Lastly, as with any research project, our backgrounds and experiences likely shaped the design and analysis of this study. Both of us worked on atomic physics experiments during our Ph.D.s, so we are familiar with similar types of apparatus, experimental techniques, and quantum concepts as those used by Oqtant. H.J.L. additionally has direct experience creating BECs in a lab. Our familiarity with atomic physics research influenced the topics we chose to include in the Oqtant educational activities, as well as our interpretations of instructor and student responses. 

\section{Results}\label{sec:results}

In this section, we present the results of this study, discussing both instructor and student perspectives of working with Oqtant for educational purposes. We begin with a sub-section about instructor perceptions, including instructors' learning goals, the challenges they anticipated facing, and the support they would need to use Oqtant in their courses. To demonstrate that these challenges are not insurmountable, we next describe one specific course implementation where Oqtant was incorporated into an upper-division quantum mechanics course. We include both the details of the implementation and the instructor's perception of how they and their students benefited from working with Oqtant. We then switch to the student perspective, describing how Oqtant felt like a real experiment to the students even though they never interacted with it physically; comparisons students made between Oqtant and other experiments they know about, including benefits and drawbacks stemming from it being remote; and the parts of working with Oqtant the students enjoyed and did not enjoy. Together, these data provide an overview of possibilities and challenges for incorporating a remote quantum experiment like Oqtant into educational settings.

\subsection{Instructor perceptions of Oqtant } \label{sec:instructorPerceptions}

Oqtant provides instructors with a potentially novel way to focus on different types of learning outcomes, yet also comes with its own set of challenges. Drawing on both the survey results of many instructors considering using Oqtant in their courses and the interview of one instructor after having incorporated Oqtant in one of their courses, we see that instructors care not only about their students learning concepts, but also about experimental skills and non-cognitive aspects. However, instructors also discuss many challenges they anticipate encountering. Some of these can be mitigated with help from outside educators through the development of educational materials for students and training materials for instructors, although there are some challenges that are inherent to working with a platform like Oqtant or would require additional support from instructors' own institutions. 

\subsubsection{Learning goals} \label{sec:learningOutcomes}

\begin{figure*}
    \centering
    \includegraphics[width=\linewidth]{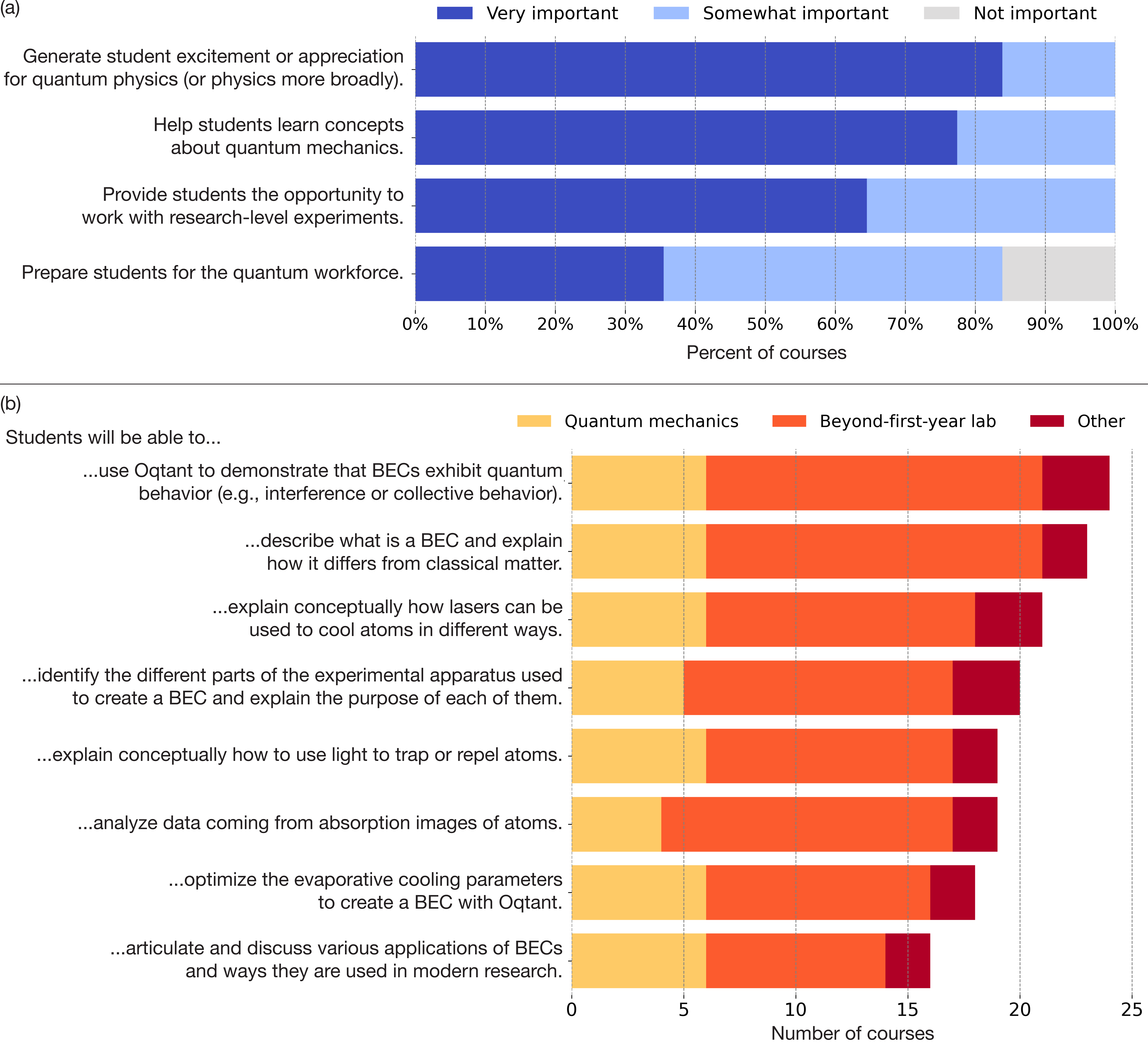}
    \caption{Learning goals instructors would value for their students if they were to use Oqtant in their courses. (a) Four broad categories of possible learning goals with the percent of courses where the instructor indicated that the listed reasons were very (dark blue), somewhat (light blue), or not (gray) important. (b) Number of courses where the instructors indicated they would be interested in, and have time to focus on, the specific learning goals in quantum mechanics (light yellow), beyond-first-year (BFY) lab (medium orange), or other (dark red) courses. The dataset contains 19 BFY lab, 8 quantum mechanics, and 4 other types of courses.}
    \label{fig:learningGoals}
\end{figure*}

Instructors have a variety of reasons for wanting to incorporate Oqtant into their courses. When asked on the survey to rank the importance of learning quantum concepts, generating excitement, working with research-level equipment, and preparing for the quantum workforce, the majority of instructors ranked all four reasons as at least somewhat important for their courses [see Fig.~\ref{fig:learningGoals}(a)]. The only reason considered not important for a small percentage of courses was preparing students for the quantum workforce. However, for the majority of the courses, instructors still regarded this as a somewhat or very important reason. An experiment like Oqtant can be used in the classroom for many different types of educational goals, and instructors often want their students to accomplish many of these concurrently.

When asked about more specific learning goals that may be possible with Oqtant, instructors were again interested in incorporating many of these into their course [see Fig.~\ref{fig:learningGoals}(b)]. For each course, instructors chose from one to eight out of eight listed learning goals plus a write-in option, and there was a mean of five learning goals per course. The learning goal that the most instructors would be interested in, and anticipate having time in their course for, was using Oqtant to demonstrate quantum behavior with BECs. However, all of the learning goals listed were chosen for between 16 and 24 courses (out of 31 in total), so all of the goals were of interest to the majority of the instructors. Instructors valued additional learning goals not mentioned on the survey as well, including understanding how theory and experiment are connected, realizing that experiments are not textbook-perfect, and developing skills such as data analysis and computation. 

Especially for instructors teaching primarily theory courses, some wanted to provide their students experiences with a ``concrete experiment'' to connect with the ``abstract theory.'' When  asked about reasons for incorporating Oqtant into their course, one instructor described the ways they wanted students to make this connection: they hoped their students would ``correlate theoretical concepts in quantum physics with experimental setups, analyze experimental results using theoretical models, and evaluate the predictions of the theory based on experimental observations.'' Another instructor emphasized the goal of providing their students an ``intro to real world quantum systems.'' Oqtant is not the only experiment that can provide students a connection to quantum experiments, but it may provide a way to do this for institutions without physical experiments or where instructors do not want their students to spend time on hands-on experimental aspects.

In addition to seeing the connection to experiments, another instructor wanted students to learn that experiments are not ideal, and that not everything can be answered by a textbook. They said,
\begin{quote}
    I like [the students] recognizing experiments aren't ideal and... thinking more broadly and deeply about why things aren't textbook... so having students sort of acknowledge and have to think about things that don't necessarily have `here's the textbook answer' was something I really wanted to do. 
\end{quote}
Oqtant provides students the opportunity to see experimental imperfections and to grapple with questions to which their instructors do not know the answer, aspects that are different than the content often covered in quantum mechanics courses. 

It is not just students who could benefit from the incorporation of Oqtant into courses, but also their instructors. Instructors can also be excited to work with a new platform and want to learn about BECs. When asked why they wanted to implement activities with Oqtant in their course, one instructor said, 
\begin{quote}
    Because I wanted to learn more about BECs... Most of the new teaching things I do is like that sounds interesting, I should know more, I'm gonna teach a class in it because that's a great way to learn. But in a more serious answer, I did want to learn a lot more about this, and it looked like a great opportunity.
\end{quote}

There are many reasons instructors consider incorporating Oqtant into their courses, and they often hope to accomplish multiple distinct learning outcomes with this experiment. These outcomes include ones typical of non-lab courses, such as conceptual learning about various topics including BECs and atom-light interactions, as well as experimental skills often emphasized in lab courses, such as data analysis. Instructors also have other broader goals including generating student excitement about physics or helping students understand the connection between theory and experiment. Although more work needs to be done to study whether or not students accomplish these learning outcomes while working with Oqtant, instructors believe there is the potential to address many of them with this platform. Oqtant may provide some of these benefits not only for the students, but for instructors as well, who may also be excited about working with a novel apparatus and learning about an area of physics that could be unfamiliar to them.

\subsubsection{Challenges} \label{sec:challenges}

Instructors also discussed a variety of concerns they have about implementing Oqtant in their courses and ways it could be challenging. The challenges identified in our thematic analysis of the instructor data are listed in Table~\ref{tab:challenges}.

\begin{table}[htbp]
\centering
\caption{Challenges instructors face or anticipate facing when incorporating Oqtant into their courses.}
\begin{tabular}{>{\hangindent=1.5em}p{8.5cm} } 
    \hline \hline
    \textbf{Challenges} \\
    \hline
    Need reliable access to Oqtant during class \\
    Instructors are not experts in topics  \\
    Students do not have necessary background knowledge\\
    Experiment different from other common experiments \\
    Could take away from hands-on experiences\\
    Need support for instructors and TAs\\
    \hline \hline
\end{tabular}
\label{tab:challenges}
\end{table}

One of the big challenges that instructors have very little control over is access to Oqtant. If instructors are planning for students to work with the experiment during a class session, they need the experiment to be online and functioning well during that time. One instructor explained this concern, saying
\begin{quote}
    If the experiment or interface was down, this could of course cause problems for students. Experimental issues arise in Advanced Laboratories all the time, but the time-honored strategies for dealing with them don't work as well from a distance.
\end{quote}
The instructor would not be able to use their usual strategies to troubleshoot Oqtant since they have no direct control over Oqtant's hardware. It could alternatively be a challenge for instructors to figure out how to help their students engage with Oqtant in class sessions during which Oqtant is not online. Instructors must also figure out how to match activities with Oqtant with the topics covered in their course when Oqtant is only online every other week.

Another challenge is that instructors are not necessarily experts in the topics needed to use and understand Oqtant, including BECs, experimental atomic physics, and programming in Python. This can make it difficult for instructors to easily convey ideas to their students, help the students when issues arise, and make course decisions related to the use of Oqtant. For example, when asked how they had graded the activities with Oqtant, the course instructor discussed how challenging it was: 
\begin{quote}
    [Grading/evaluating the activities] was the hardest thing. By far. Because... especially with a topic like this where it's not the main center of my research, I think I understand things at an appropriate level and get it, but I don't necessarily know... if what I think is important to be covered is what they do. So I think developing clear rubrics and all that... was a challenge for me.
\end{quote}
It was also a challenge to figure out when in the course to implement the activities with Oqtant so they fit with the content of the course, since the instructor did not have a complete understanding of the underlying material when finalizing the syllabus.

The lack of familiarity with BECs and experimental atomic physics is also relevant for the students, since these topics are only minimally covered in most undergraduate physics curricula. One instructor discussed their concern that they were ``not sure which material would be useful, enough to introduce the subject but not excessive.'' This is especially true for the instructors who were considering incorporating Oqtant into a BFY lab course where their students would not necessarily have already taken courses in quantum mechanics or statistical mechanics. It is an open question whether Oqtant could be successfully incorporated into courses where students did not have a prior basic understanding of the underlying content.

Additionally, the apparatus used in Oqtant is different than other experiments students have worked with before, which adds another set of elements the students must learn during the course. When asked what they would be concerned about when incorporating Oqtant into their course, one instructor responded,
\begin{quote}
    The stark difference in apparatus from most of the other course's experiments. There tends to be a high cognitive overload for students due to having to learn to use new apparatus for many experiments.
\end{quote}
Another instructor also saw this as concerning, not necessarily because it would be too much for the students, but because the new learning would not allow for any buffer time if things went wrong. This instructor explained,
\begin{quote}
    My only concern is the amount of `overhead' required to get the system running and familiarize the students with the system. There will not be time to go back if something fails during the time we would be able to dedicate to this activity.
\end{quote}
Learning how to work with a new type of experiment, especially one as complex as Oqtant, requires both time and cognitive load. 

Instructors were also concerned that using Oqtant would take away time students could spend working with hands-on experiments. One instructor stated how they would only use a remote experiment if in-person labs were not an option:
\begin{quote}
    Current students desperately need actual hands-on experience with experiments -- connecting cables, turning knobs, etc. Even when I introduce computer control of an experiment on the table in front of them, I see students missing the fact that they need to look at / think about the actual direction the wave plates turn in reality, not just the numbers on the screen. In a lab course or lab portion of a course, I would pretty much always choose to have students do an actual, in-person experiment (even if it has to be simpler and less impressive) over a virtual or web-based one. I use virtual or web-based `experiments' like this in non-lab courses only, or as outside-lab-time training exercises to go with in-person labs in a lab course.
\end{quote}
Other instructors would consider using a remote experiment as long as their students were able to debug some aspects of it and have control over parts of the experiment. For one instructor, part of the challenge was convincing other faculty of the value of this type of remote experiment:
\begin{quote}
    The hands-on aspect of the laboratory is an important component of the course. It is hard to engage other faculty on the value of cloud-based experimental work... I think a significant component of the laboratory activity should include some debugging.
\end{quote}

Finally, instructors were also concerned about the need for training of instructors and teaching assistants (TAs) so that they would be able to quickly and easily help students while working with Oqtant. More details about the kind of support instructors would need to address this and the other challenges are included in the following sub-section.

\subsubsection{Support needed}

Some of the challenges instructors discussed could be eased through various types of support. The support instructors requested when asked what they would need to feel comfortable incorporating Oqtant into their course are listed in Table~\ref{tab:support}.

\begin{table}[htbp]
\centering
\caption{Themes from the thematic analysis identifying types of support instructors would need to implement Oqtant in their courses.}
\begin{tabular}{>{\hangindent=1.5em}p{8.5cm} } 
    \hline \hline
    \textbf{Support needed} \\
    \hline
    Educational materials for students \\
    Background materials for instructors \\
    Training for instructors and TAs \\
    Time to work through activities and learn about Oqtant\\
    A point of contact to answer questions and troubleshoot \\
    Platform for educators to share resources\\
    \hline \hline
\end{tabular}
\label{tab:support}
\end{table}

Instructors discussed the need for various types of educational materials for their students. This includes lab guides or activities, preparation activities for beforehand, and example homework problems. Some instructors emphasized that these materials should allow opportunities for student decision-making and exploration. For example, one instructor said:
\begin{quote}
    We would probably need help creating a lab manual for the students that properly walks them through the experiment at the appropriate level that leaves space for inquiry and exploration, but allows [them] to stay within the parameters of the experiment.
\end{quote}
Another requested
\begin{quote}
    A set of questions/hypotheses that can be answered with the [Oqtant] system or a framework for how students could engage meaningfully with and understand/test the limitations of the system they are working with.
\end{quote}
When experiments are as complex as Oqtant, lab guides often end up being relatively guided; however, open-ended aspects of experiments provide students many benefits \cite{wilcox2016open, smith2021best, liu2024correlation}. Oqtant has the possibility to provide students space to explore, and instructors wanted guidance about manageable ways to do that.

Instructors also wanted materials to help themselves learn about BECs and the apparatus, as well as other training opportunities. One instructor said, ``The best thing for me is to have sufficient materials and documentation and a straightforward access process so I can be as self-reliant as possible.'' In addition to background materials and good documentation, instructors requested instructor guides with sample solutions and guidelines on the timing of the activities. Instructors also asked for various kinds of training that could include webinars, courses, or workshops. At some institutions, there are different instructors and TAs involved in the relevant course each year, so the training and support would need to be ongoing. 

Along with the provision of sufficient materials, instructors would need time to work through the activities and learn about the experimental platform on their own. The instructor who implemented Oqtant in their course explained how they would have liked an easier way to learn the material since they did not have a lot of time to spend reading references:
\begin{quote}
    I mentioned that one colleague, the lecturer who did a PhD in atomic physics... and I remember pinging him, and immediately `here's my grad level text.' And I was just like... I just need a draw-me-a-picture sort of thing because I don't have the bandwidth to do a full deep dive... it's not a lack or failure of resources or anything. It's just time I didn't have to put into it that I probably should have.
\end{quote}
This instructor points out that useful support could include either more time (which is often not feasible given many instructors' workloads) or available references at a different level. 

Instructors also discussed the importance of various methods of communication with other educators and Oqtant staff. They wanted a point of contact (presumably either the educators who developed the activities or employees at Infleqtion who work with Oqtant) to help answer questions or troubleshoot any issues that arise. Instructors also discussed the idea of a platform where they could ``share resources and discuss their experiences'' with other instructors using Oqtant in their classes. It is not clear whether this would be best facilitated by the staff working on Oqtant (e.g., as part of the Oqtant Slack channel) or through some other forum operated by educators.

Overall, instructors anticipated or experienced various challenges to working with Oqtant, some of which could be addressed with the various methods of support discussed in this section. Some of these support mechanisms (such as the provision of educational materials and training opportunities) could be provided by outside educators (i.e., other instructors or education researchers). However, substantial resources are needed to sustainably develop and maintain such materials and opportunities for professional development \cite{barnes2024outcomes}.

\subsection{Implementation of Oqtant in a course}

In order to help instructors understand how these challenges can be managed and what learning outcomes they might expect when Oqtant is implemented in a course, we present here a concrete example of Oqtant being used in an upper-division quantum mechanics course. This illustrates the way an instructor, who is not an expert in experimental atomic physics or BECs, was able to fit two structured activities and a final project with Oqtant into their course, as well as how they navigated various practical challenges, such as the way their course time did not match up with Oqtant's online hours. Overall, the instructor enjoyed implementing Oqtant in their course, thought their students met their learning goals related to engaging with authentic scientific experiences, and would like to implement Oqtant in the course again in the future.

\subsubsection{Details of implementation} \label{sec:implementation}

The instructor we partnered with chose to implement Oqtant in a second-semester quantum mechanics course for upper-division physics students. The course was ``one of the terminal courses'' in the physics major at a primarily-undergraduate institution and covered topics such as the quantum harmonic oscillator; hydrogen atom; perturbation theory; hyperfine, fine structure, and Zeeman Effect; and identical particles. There were a total of 12 students in the class, consisting of mostly seniors in their last or second-to-last semester of college. The class met once a week for 2 hours 50 minutes each Friday throughout the Spring 2024 semester. Most of the students had previously taken a computational physics course in the department, although that course did not use Python.

The instructor did not have expertise in some areas relevant to Oqtant, although they had taught this quantum mechanics course multiple times before. The instructor's research area is experimental solid state physics, and they reported having very little knowledge about experimental atomic physics or BECs when deciding to implement Oqtant in their course. They had some experience coding and described their level of knowledge of Python as ``functional familiarity.'' They had taught this quantum mechanics course 2-3 times before, and each time they had kept the course mostly the same, but varied the final presentations or papers.

This instructor implemented the two activities we developed at various points throughout the semester and ended the course with an open-ended project with Oqtant. The instructor picked two locations in the course and inserted these activities at a time when the topic changed (e.g., from the hydrogen atom to perturbation theory), so the activities with Oqtant did not break up the flow of the rest of the class. The instructor assigned the preparation activities, had students submit their responses before class, and treated them the same as other pre-lecture questions where students could discuss them briefly at the start of the class. Oqtant was never online during the class sessions, so the students worked on Oqtant activities together in class, but the jobs they submitted were only run outside of class time. The details of how each activity was included in the course is depicted in Table~\ref{tab:courseSchedule}. The instructor asked us or Infleqtion employees working with Oqtant questions a few times throughout the semester when the online documentation was not sufficient.

\begin{table*}[htbp]
\renewcommand{\arraystretch}{1.4}
\centering
\caption{How Oqtant was used in an upper-division quantum mechanics course by week. Oqtant online means the experimental apparatus was running the three days prior to the class sessions (which occurred on Fridays). Although Oqtant was online approximately every other week, there were some schedule irregularities. During the course, students completed two Activities (A), each preceded by a Preparation Activities (PA), as well as an open-ended final project that culminated in a presentation (pres.) to the entire class. For the main activities, the students wrote code to submit jobs and analyzed data from Oqtant in class, and their jobs were run outside of class time.}
\begin{tabular}{P{1cm} P{0.5cm} P{0.5cm} P{0.5cm} P{0.5cm} P{0.5cm} P{1.55cm} P{1.25cm} P{0.5cm} P{1.25cm} P{0.5cm} P{0.5cm} P{1.55cm} P{1.25cm} P{1.25cm} P{1.25cm} P{1.25cm}} 
    \hline \hline
    & \multicolumn{15}{c}{Week of Semester} \\
    & 1 & 2 & 3 & 4 & 5 & 6 & 7 & 8 & 9 & 10 & 11 & 12 & 13 & 14 & 15 & 16 \\
    \hline
    Oqtant online & & \checkmark && \checkmark && \checkmark & \checkmark & \checkmark && \checkmark &&& \checkmark && \checkmark & \\
    Out of class &&&&&& PA1 & Jobs for A1 run && &&& PA2 & Jobs for A2 run & Plan project & Submit jobs for project & Prepare pres. \\
    In class &&&&&& A1 data analysis and job submission & A1 debrief && (Spring break) &&& A2 data analysis and job submission & A2 data analysis &&& Project pres. \\
    \hline \hline
\end{tabular}
\label{tab:courseSchedule}
\end{table*}

The students worked through the majority of the first activity in a single class session. The first 20 minutes of that class session were spent discussing the preparation activity, and then the students worked through the main activity for the rest of the class while the instructor circled around, helping students as needed. That activity began with the students analyzing already taken data, so it was not a problem for most of the activity that Oqtant was not online at the time.  However, the end of the activity required students to submit their own jobs to Oqtant, so students submitted jobs to the job queue by the end of the class session, even though their jobs did not run on the hardware until the following week. The instructor then spent a little time discussing this activity the following class session after the students were able to collect the data from their jobs. The instructor intended to have the students work in small groups for this activity, but, as was typical in the course, the students ended up discussing the activity as an entire class.

The second activity was incorporated into the class over two sessions, surrounding a week when Oqtant was online. In total, the instructor dedicated around 4--4.5 hours of class time to this activity. During the first week, the students performed some new analyses of the data they had taken in Activity 1 and planned out what new jobs they were going to request and submitted those. They then retrieved the completed jobs and finished the activity the second week. The students worked in their final projects groups for this activity so they could begin generating ideas for their final projects, and there was additionally communication between groups.

The final project was a way for students to explore topics they were interested in with Oqtant in small groups at the end of the semester. The instructor encouraged the students to choose a topic that the class had ``swept under the rug or something that piqued [their] interest.'' All four of the three-student groups ended up starting with one of the demonstration notebooks Oqtant hosts online \cite{oqtantDemoNotebooks} focusing on optical barriers and tunneling, quantum interference, and the speed of sound in quantum matter. No in-class time was dedicated to working on the final project, but the students had 2--3 weeks to complete it, with Oqtant being online one of those weeks.

To grade the activities with Oqtant, the instructor developed rubrics. For the preparation activities, the instructor tried to check whether the students' answers lined up with their own understanding and if the student reasoning ``show[ed] some deep thought.'' For the activities that the students submitted as Jupyter notebooks, the instructor applied their rubric to the notebooks without running any of the cells, and gave a set number of points for certain analyses. The activities and final project with Oqtant combined to be worth up to a total of 15\% of students' grades.

\subsubsection{Perceived outcomes} \label{sec:instructorOutcomes}

The instructor thought that the students in their class had accomplished the learning goals of thinking critically about experimental data and learning how to work with it. When asked if their learning goals were met, the instructor said they were ``happy with how it went'' although they acknowledged that some students ``did it better than others.'' When discussing skills the students had gained, the instructor further explained that ``the major thing [the students] got out of'' working with Oqtant was ``critical thinking about experiments and data analysis,'' which included a ``better understanding of... fitting data to functions to extract physical parameters.'' However, these skills may not have come just from working with Oqtant as many students in the course were concurrently enrolled in an Advanced Lab course where those data analysis skills were also practiced.

The instructor also emphasized how engaging with Oqtant allowed their students to work on problems that were not clearly defined and experience the messiness of science. When asked about other benefits working with Oqtant provided to their students, they said, 
\begin{quote}
    It provided a good opportunity for them to... get comfortable with true science experiences... They saw me, as someone who hopefully they perceive as a scientist or a physicist, go through probably similar struggles they had where it's like, `I really don't know the answer here, but I'm comfortable with that, and we can figure it out'... I think that adds value for students to feel more comfortable... 
\end{quote}
It can be valuable for students to see examples of their instructor not knowing all of the answers, as this is how students engage with experts in research. The instructor also wanted students to see how experiments do not always follow a linear trajectory:
\begin{quote}
    One of the things that I think is really hard to get at in the curriculum is the messiness of actual science... And just that realistic approach to experiment, I think, is a valuable thing that it was nice to be able to have a nice sort of experience that sort of walked students through that.
\end{quote}
They saw working with Oqtant as a way to guide students through an authentic type of science experience.

The instructor thought the iterative approach required by Oqtant also provided students an opportunity to learn about some aspects of research projects.  When discussing the skills their students had gained, the instructor brought up the idea of course-based undergraduate research experiences (CUREs) \cite{auchincloss2014assessment, buchanan2022current, merritt2024physics}. The instructor discussed how this experience with Oqtant may have given students a similar view of what science is like to what they might learn in a CURE. They said,
\begin{quote}
    I think that they got a similar type experience, or what I expect [CUREs] would be like... because a lot of those final projects... were like, `yeah, so this clearly was not what we expected sort of thing.'
\end{quote}
They continued to explain how seeing unexpected results prompted the students to consider various reasons for the discrepancy, a skill the instructor values in students in their research group.
Incorporating Oqtant into the course provided the students an opportunity to see how experiments rarely work the first time and to therefore help the students engage in the authentic practice of iteration \cite{dounas2016nothing}.

Although the open-ended nature of the final projects helped provide students the opportunity to engage in some authentic scientific practices, the students may not have been fully comfortable with the ambiguity that came with it. The instructor explained, 
\begin{quote}
    Quite a few groups were concerned... that it was kind of an open-ended project. So they wanted some more concrete like is `this a good question to answer,' and-- They didn't really appreciate the `I want you to have learned something and then teach it to me and convince me you've learned it and whatever interests you like, there's no bad question here.'... I don't think they were comfortable with that open-ended... type question.
\end{quote}
This lack of comfort with open-ended assignments is not unique to Oqtant; the uncertainty and the lack of structure and guidance has been shown to make some students feel uncomfortable \cite{henige2011undergraduate,owens2020student}, while others appreciate the freedom associated with the open-ended projects \cite{henige2011undergraduate,kalender2021restructuring}.

Some of the students seemed to really enjoy working with Oqtant and got a lot out of it, whereas others may have thought it was not worth the  time and effort they needed to dedicate to it. When asked how students perceived working with Oqtant, the instructor said,
\begin{quote}
    I have some sampling bias, because I've only heard back... from the ones who really enjoyed it... So I've had a couple students mention how much they enjoyed it and really liked it. One student... said `if [my summer research plan] all falls through and I don't have anything going on in the summer, I would love to keep working with Oqtant and maybe try to... see if [I] can put together notebooks for the Infleqtion site or something'... so two or three I think really liked it... I expect there were the other tail as well where two or three were probably like this was a lot of work and not worth my time.
\end{quote}

This perceived mixed response from the students carried over to conceptual understanding of Oqtant as well. When asked if the activities were at the appropriate level for their students, the instructor said,
\begin{quote}
    I think so. I think some of the students who did better with it in my class really understood a lot of it. Some of the students who succeeded, but I wouldn't say excelled... had discussions with me a bit like, `alright, I got it to work and we did it, but I'm having a hard time thinking about the physics because it took all my effort to do that.'
\end{quote}
There were a lot of concepts (both physics and Python coding) that all needed to connect together for students to understand what was happening in the experiment, so it is unsurprising this may have been a challenge for some students.

The instructor also enjoyed implementing Oqtant and learned from it. When asked if they had enjoyed incorporating Oqtant into their course, they said,
\begin{quote}
    I did. I like to learn things. And that's kind of one of the motivations for doing it was I wanted to actually learn more than `oh yeah, BECs and things are cool.' And so I got to play with it. And even though I just finished saying I didn't take as deep a dive [into the underlying theory] as I would have liked, I did learn a lot. And it was fun. I had a lot of fun working with the students and watching them learn.
\end{quote}

Overall, the instructor thought the implementation of Oqtant was a success. Although some students enjoyed working with Oqtant more than others, the students engaged in a scientific experience that was more authentic than the activities in many of their other courses. In particular, the way the instructor did not have all the answers and the iterative data collection and analysis process helped the students think critically about experimental data and learn about the scientific process. Because of this success, the instructor is planning to incorporate the activities and final project with Oqtant into their course again with only minor changes, such as spending slightly more time discussing the preparation activities.

\subsection{Student perceptions of using Oqtant}

In the previous section, we saw that the instructor of the course thought some of their students enjoyed working with Oqtant and gained skills from doing so, but it is important to investigate the students' perceptions of their experiences too. By looking at students' written reflections after the two structured activities with Oqtant, as well as the think-aloud interviews, we find that students do perceive Oqtant to be a real experiment even though they never interact with the apparatus in-person. Additionally, they think that it can provide opportunities for them to experience a type of quantum experiment that may not be available at their home institution, although the remote nature of the experiment comes with some drawbacks as well. Some of the students enjoyed being able to connect to an experiment in a different location, although they did not like some of the necessary infrastructure that came with it. 

\subsubsection{`Realness' of experiment} \label{sec:results_real}

One question that naturally arises when students do not have physical access to an experiment, is whether or not students perceive themselves to be working with a real experiment. When we directly asked the students, most of them agreed that working with Oqtant did feel like working with a real experiment even though they could only access it remotely. They cited a variety of reasons for this, which are listed in Table~\ref{tab:real}.

\begin{table}[htbp]
\centering
\caption{Reasons why students felt like they were working with a real experiment as they were working with Oqtant, as identified in the thematic analysis of the student data.}
\begin{tabular}{>{\hangindent=1.5em}p{8.5cm} } 
    \hline \hline
    Students felt like Oqtant was a real experiment because... \\
    \hspace{.2cm} ...they worked with real data they had requested. \\
    \hspace{.2cm} ...they chose the experimental parameters. \\
    \hspace{.2cm} ...there was variation and noise in the data. \\
    \hspace{.2cm} ...there were experimental imperfections.\\
    \hspace{.2cm} ...they needed to interpret the data. \\
    \hspace{.2cm} ...they were able to troubleshoot the experiment. \\
    \hspace{.2cm} ...they had learned how the apparatus functions. \\
    \hline \hline
\end{tabular}
\label{tab:real}
\end{table}

Oqtant seemed like a real experiment to some students because they worked with real data that they had requested. One student said ``Even though I could only access it remotely, I felt that when I submitted jobs I was actively requesting data from a lab I know exists but physically can't visit myself.'' Other students emphasized the choice they had in the experimental parameters they requested. One student explained, ``If I can change certain parameters to see how a system is affected, that is a `real' experiment. I believe that this would be essentially the same if the technology was at our university campus or at home.'' Another student agreed with this idea by stating ``it feels like there is a lot of depth here in terms of what parameters you have control over... that helps it feel more like an experiment rather than some cloud-based thing that I'm just running code on.''

Even when students choose the same parameters, the way the returned data was not always the same, and it had noise and variation, also contributed to Oqtant seeming like it was a real experiment. One student said, ``the labs seem like we are working with a real experiment because we have the choice of what parameters we want, and even if we choose the same ones, the data isn't the same.'' Another student explained, ``it made me feel like I was actually getting data because I was inputting a job, and then there was actual noise and variation at the output of it.'' 

Some students noticed other experimental imperfections that would not have occurred if it were not a real experiment. One student who had noted a speck of dust on a returned image said ``I guess the speck of dust was cool. It made me realize that it's an actual experiment too.'' Another commented not on imperfections of the returned data, but on the way the experiment schedule was not ideal, saying ``This activity did make me feel like I was working on a real experiment, especially considering the downtimes the lab had for updates and other processes.''

Connected with several of these ideas is the way that students needed to interpret the data they obtained from Oqtant. One student said,
\begin{quote}
    I actually had to interpret it... When it comes to like determining where the BEC was created, there's some variability of where I can choose, and the temperature and the ranges... That actually felt like I had to do some thinking about it, and analyze the data itself and actually work from it.
\end{quote}
Another student compared their analysis of data from Oqtant with a simulation, explaining, ``we get to analyze much more realistic data that has flaws, forcing us to more deeply analyze and interpret the raw data to reach conclusions instead of being spoon fed the answers like we are with idealistic data.'' Other ways of needing to think about the experiment also contributed to students feeling like it was a real experiment including the ability to troubleshoot when issues arose and the way the students understood how the apparatus functions.

On the other hand, there were reasons why Oqtant may not have felt like a real experiment to students. One student said, ``The only thing that kind of felt like it could be something like a simulation was the fact that it was completely remote and the idea that someone could potentially create a simulation with imperfections.''
They followed up on this idea, asking
\begin{quote}
    But, how would I know [if it was a simulation]? Like, if I'm just some person, just kind of messing around with this? I'm just curious. I mean, because it's all through a [Jupyter] notebook. And it's all remote. You could just make a simulation that models BECs and people will have no idea. You know?
\end{quote}

Overall, most students perceived Oqtant to be a real experiment for a variety of reasons. Some of these reasons are inherent to experimental physics (e.g.,  noise in images returned by the experiment), whereas others were based off of actions the students took (e.g., choosing the parameters, requesting data, and interpreting it). Instructors could consider discussing these different elements with their students, emphasizing the capabilities and limitations of real experiments.

\subsubsection{Comparison with other experiments} \label{sec:comparison}

Although Oqtant is a new type of publicly available experimental platform, there are many ways it is both similar to and different from other physics experiments. We asked students to compare Oqtant with other experiments they know about in both instructional and research labs, noting that they came in to this study with varied prior experiences. Students pointed out various similarities and differences, focusing both on the experiment itself (e.g., the system it studies, type of data, and data analysis needed) and some aspects of being remote (e.g., easy access and controlling the experiment only from a computer). Students additionally discussed how the remote nature of Oqtant allows for potential benefits over non-remote experiments, while also coming with some drawbacks. The characteristics of Oqtant students discussed, along with ways different students associated them, are listed in Table~\ref{tab:comparison_v2}. 

\begin{table*}[htbp]
\centering
\caption{Comparison between Oqtant and other experiments students had worked with (or knew about) in instructional or research labs. The check marks indicate whether these emergent characteristics were discussed as being similar to or different from students' experiences, as well as whether they were related to noted benefits or drawbacks of remote experiments. Multiple check marks for seemingly contradictory classifications indicate that different students talked about the characteristics in different ways, and the lack of a check mark indicates that the connection was not discussed unprompted by any of the students.}
\begin{tabular}{l c c c c} 
    \hline \hline
    \textbf{Characteristics of Oqtant} & \textbf{Similar} & \textbf{Different} & \textbf{Benefit} & \textbf{Drawback} \\
    \hline
    Ability to take and analyze data & \checkmark & & &  \\
    Data analysis involves plotting and fitting data & \checkmark & \checkmark & & \\
    New type of physical system (atoms and optics) & \checkmark & \checkmark & \checkmark & \checkmark \\
    Anyone can access experiment from anywhere & & \checkmark & \checkmark & \\
    Quick and easy way to obtain a lot of real experimental data& & \checkmark & \checkmark & \\
    Cannot physically see or interact with setup & & \checkmark & & \checkmark \\
    Control experiment from computer interface & \checkmark & & \checkmark & \checkmark \\
    Remotely connecting to experiment operated by someone else &\checkmark & \checkmark & \checkmark & \checkmark \\
    \hline \hline
\end{tabular}
\label{tab:comparison_v2}
\end{table*}

The students found the general experimental process with Oqtant of taking and analyzing data to be similar to their other experimental experiences. When asked how the data analysis they performed in the first activity compared with their prior labs, one student said, ``This lab shares many similarities to previous labs [in courses] in the sense that there was data to be obtained through a streamlined process.'' Another student compared Oqtant with other apparatus in research labs, saying, ``I would say it's similar: you take data and analyze the data to fit different models, similar to other research labs.'' 

However, the specific type of data analysis students performed with Oqtant was similar to some students' prior experiences yet different from others'. When comparing the data analysis from the first activity with their prior lab courses and research experiences, one student said, ``The data analysis here was pretty similar to what I've done in the past... In many of my plots, I have to fit data points with a nonlinear equation, and compare results between two or more plots, just like in this lab.'' Other students pointed out some differences including the specific fit functions (e.g., a 2D Gaussian) and the quantitative nature of the analysis. One student appreciated the opportunity to use a Gaussian fit since they had heard a lot about that fit function without having an opportunity to apply it.

Not only the type of data analysis, but also the physical system under study was familiar for some students, but not others. When comparing Oqtant with other apparatus in research labs, one student explained the similarities: ``This Oqtant apparatus is definitely a specialized piece of equipment; however, it is not vastly different from many optical labs as it uses systems that can be used in other fields of physics.'' Another student pointed out that other research labs also ``play with BECs.'' Some students found Oqtant different from their prior experiences. One student, who worked on ``more of the computing part and the theory part'' explained how working with atoms was new for them: ``So I guess it's just a more physics-y part of quantum that I haven't really explored, that isn't really taught here at [my institution].'' They explained how Oqtant's emphasis on the experimental hardware and ``actual measurements'' (as opposed to creating algorithms) was a new opportunity for them. However, not all students appreciated working with a new type of system. One student brought up their lack of experience with the system as a drawback, although the same would also be true for any other new type of experiment.

The way this type of system is made publicly available and can be accessed from anywhere is something students noted as being both unique and a large benefit. One student explained, ``working on a remote experiment is great because I don't need to travel to a different state in order to do experiments, and others can run the same experiment with just an internet connection.'' For some students, this was a way to bring a cold atoms experiment to their institution, since they did not have access to such an experiment previously. One student said, 
\begin{quote}
    It... reduces the lack of professional information I have access to. Because that's one of the big issues at [my institution]. We don't have a lot of quantum teachers yet... So I think remote experiments gives me access to those different informations.
\end{quote}
This student saw the benefit for themselves that educators are hoping Oqtant could broadly provide: access to quantum hardware for students who would not otherwise have the opportunity \cite{tingle2024oqtant}. 

Due to the easy access, Oqtant provides students a simple and quick way to obtain a lot of real experimental data. This allows them to work with data from a complex experiment without needing the time, equipment, and technical knowledge to build it themselves. The ease of obtaining data allows students to collect a lot more data from Oqtant than they do from other experiments, and the data contains real experimental noise and variation compared with what they could obtain in a similarly easy manner from a simulation. One student explained how Oqtant also works better than other apparatus in research labs, saying ``it seems to always work as intended, which I think is super cool.'' Students are therefore able to focus on the data collection and analysis aspects of experimentation as opposed to building and maintaining the apparatus. One student explained,
\begin{quote}
    I find that Oqtant is a hands-off type of experimentation that allows for researchers to make use of the equipment and operate it, without having the hassle of operating and maintaining it. This is very convenient, because it allows researchers to focus more on conceptual ideas, as opposed to experimental setup.
\end{quote}
Oqtant may make it easier for students to engage with certain aspects of experimental physics, even though it does not allow for others.

Although the students noted important benefits to working with an experiment like Oqtant, one of the drawbacks to being remote, and also a difference from most students' prior experiences, is that they could not physically see or interact with the experimental setup. One student explained how different this was for them: ``This definitely feels a lot different from a traditional lab course where it would be a lot more hands-on with... setting up the optics and doing the alignment.'' Some students thought the inability to see the apparatus detracted slightly from their experience, saying that there is ``a little bit less intuition behind what your goal is'' and that it makes it less ``tangible.'' Nonetheless, students can quickly become used to not seeing the physical apparatus. One student said,
\begin{quote}
    At first, it's hard to get used to, but when you're used to it... it still gives you everything you need. So, I think since this is the second time I've done this, a lot more used to it. And so I don't think it really affected my experience that much.
\end{quote}
Students have talked about the importance of ``seeing'' the experiment themselves in other contexts too, such as with in-person quantum optics experiments \cite{borish2023seeing}.

Instead of interacting with the experiment physically, students controlled the experiment via a computer interface, something that was similar for some, but not all, students and that led to both benefits and drawbacks. Some students had prior experience with or knowledge about IBM's quantum computer \cite{IBMquantumComputing} and discussed similarities of Oqtant with that, especially how both experiments involve users writing code that they run by submitting a limited number of jobs per day. When comparing Oqtant with other research labs, one student said, ``the majority of my experiences when working with quantum devices is usually interfaced similarly with jobs / only accessing some terminal or GUI of the actual device.'' It was not clear if this student was talking about working with IBM's quantum computer or experiments in research labs they had worked in, since many experiments in non-remote labs are run from a computer. They further explained,
\begin{quote}
    Me having experience in the lab, a lot of the time it is just, `oh, this is what it is, and now we control it from the computer.' So it feels the same, at least to me. Like once you set it up, you're just using it based off queueing commands on the computer.
\end{quote}

The fact that students could interact with Oqtant only via a computer led to opportunities for skill development as well as challenges. One student said that this experience helped them ``practice syntax and working with python coding in a designated environment.'' Although that is something students can do with in-person experiments as well, it is not possible to operate a remote experiment without interacting with some type of computer interface. However, students also noted various issues that arose when connecting to Oqtant, including bad internet connections, lack of knowledge about APIs, and trouble ensuring they had the correct version of Python. Getting Oqtant's Python package downloaded on all students' computers was a challenge also identified by the instructor of the course and us during the think-aloud interviews.

The last aspect of remote experiments students discussed is how they worked with an experiment that was operated by other people. This was a familiar experience for students who had used IBM's quantum computer, but very different for others, and it also led to both benefits and drawbacks. Although the students themselves were not directly interacting with the staff at Infleqtion, one student mentioned that collaboration is a benefit of remote experiments. They said, ``I got to... submit work requests to remote researchers, therefore practicing collaboration.'' This type of collaboration is not common in most lab courses, but is typical in some research fields (e.g., astronomy or high-energy physics). Obtaining data from an apparatus that others set up and maintain requires placing trust in people the students do not know, something one student noted as a potential problem. They said, ``I also have to place a certain level of trust that the results I get back are accurate without any way to test that.'' Although the students saw this as a drawback, it is similar to many research settings, so instructors could emphasize steps students can take to ensure they trust their data as a way to build this authentic scientific skill.

The fact that they were taking data on someone else's experiment also made it harder for students to work on their own schedule, troubleshoot the experiment, and obtain the needed support. This may be the first time some students were not able to take experimental data whenever they wanted, as they could only obtain data when Oqtant was online, and even then they did not receive the data instantaneously because of the queue. It was also hard for students to troubleshoot an apparatus that they were not familiar with. One student pointed out that even if they were around it in person, they might not be able to help troubleshoot it: ``I wouldn't say I couldn't troubleshoot the machine, but I don't think I would have been able to do anything if I had access to it. I'd just kind of watch them fix it or turn it on.'' Troubleshooting is an important lab skill for students to learn \cite{kozminski2014aapt,dounas2017electronics}, but troubleshooting experimental apparatus may be difficult to practice with a remote experiment, although there are other types of troubleshooting students could still engage with. Students also discussed how it can be hard to ask questions and get support with a remote experiment. 

Students identified many similarities and differences between Oqtant and other experiments, which they saw in both positive and negative ways. These comparisons centered around the type of experiment Oqtant is, as well as its remote nature. Some students benefited from access to a modern quantum experiment not available at their institution, one of the primary goals of using Oqtant for education. The students also noted other benefits including the ability to easily access a lot of experimental data and the opportunity to practice different skills. The drawbacks of remote experiments students mentioned include ones that are inherent to experiments like Oqtant, such as the way students cannot physically see the apparatus or they may run into issues connecting to it from a computer. However, others drawbacks noted by students (e.g., the fact that they need to rely on others to get data) could be re-framed in terms of opportunities to teach authentic scientific skills. When teaching with this new experimental platform, instructors can decide to emphasize points of common similarity to students' past experiences to increase student comfort or instead emphasize the similarity to some research experiments as a way to teach scientific skills and practices that are often not focused on in hands-on lab courses.

\subsubsection{Parts enjoyed} \label{sec:enjoyment}

Generating student enjoyment and motivation is a common goal for instructors when incorporating quantum experiments into their courses (e.g., see Sec.~\ref{sec:learningOutcomes} and Ref.~\cite{borish2023implementation}), so we investigated what parts of working with Oqtant the students particularly enjoyed or did not enjoy. Table~\ref{tab:enjoy} shows the identified themes related to student enjoyment. We focus on the aspects that are not specific to the individual activities the students completed, but could also relate to other remote experiments. 

\begin{table*}[htbp]
\centering
\caption{The parts of working with Oqtant students enjoyed and did not enjoy.}
\begin{tabular}{>{\hangindent=1em}>{\raggedright}p{5.7cm} >{\hangindent=1em}p{5.5cm}} 
    \hline \hline
    \textbf{What students enjoyed} & \textbf{What they did not enjoy} \\
    \hline
    Working with ultracold atoms & Job queues \\
    Having visual data& Python package installation issues \\
    Taking real experimental data & Not enough background information \\
    Connecting to a far-away experiment & Things not working \\
    Possibilities to explore on own  & \\
    \hline \hline
\end{tabular}
\label{tab:enjoy}
\end{table*}

Students enjoyed working with and detecting ultracold atoms, an experimental platform that was distinct from what some students had previous experience with (see Sec.~\ref{sec:comparison}). When asked what parts (if any) of the second activity they had enjoyed, one student said,
\begin{quote}
    Being able to measure the nano-Kelvin regime is really cool... And then being able to basically turn atoms into a different state and find out if that actually happened. Of course, like the process of cooling the atoms to this temperature is really interesting. Lots of really interesting physics there.    
\end{quote}
This student appreciated the physics happening in a regime they are not often able to interact with. Another student enjoyed the way the raw data came in the form of images of the atoms, saying,
\begin{quote}
    I liked... being able to see the condensate... And I liked having the data as images as well... I'm such a visual person, I have to be able to see things. And so that was really helpful. Because... when I was loading in... this data array, I was like, I actually have no idea what I'm going to be looking at, because these are just numbers... the visualization was really good for something that's hard to visualize. So I appreciated that.
\end{quote}

Students also enjoyed being able to take the data themselves. When discussing the final part of the second activity, one student said, ``even though we're seeing it over a computer screen, these runs still seem like `my' personal runs which makes it that much more fun.'' When discussing what parts of the first activity they had enjoyed, another student said,
\begin{quote}
    It's really cool... because I was actually doing experiments with whatever machine you guys have in the nano Kelvin regime. I think that's just really cool. Just by like just sending stuff... I feel like I don't deserve all the data that's coming out from that. Because I've just sent like a couple commands over. I think that's really cool.
\end{quote}
Some students enjoyed the fact that they were able to connect to an experiment in a different location.

For other students it was not just the fact that they collected data, but that they also had the ability to explore the experiment on their own. When asked what parts of the second activity they enjoyed, one student said,
\begin{quote}
    I think it's really cool how much control over these parameters you have. And I think it could be a lot of fun to just play around with... this more. And like, in lab one, I remember... we were making BECs and I was choosing the temperature. And I was like, well, I want to see what happens if I choose zero nanokelvins. And it was cool that I could do that... There wasn't anything stopping me... It seems like there aren't too many experimental barriers here. Like the space in which I can play around with this is pretty open. So I enjoyed that aspect of it.
\end{quote}

Although some students indicated there were no particular aspects of working with Oqtant they did not enjoy, there were a few aspects other students disliked including the job queues and issues with the Python package installation. Some of the students mentioned how working with Oqtant would have been more challenging and potentially frustrating for them if they were less proficient in Python. It is possible that student enjoyment, at least when accessing Oqtant through the Python API, may depend somewhat on their Python ability.

Some students did not enjoy the fact that they did not have enough background information or context when working with Oqtant. When asked about parts they did not enjoy, one student said,
\begin{quote}
    I guess the only thing is... if there were more resources. Well, I guess... there were references that I could read and stuff in the preliminary stuff... if I had a little bit more context about why we would want to measure the number of atoms and the temperature.
\end{quote}
Other students also mentioned how the background reading was dense or they did not retain all of it. Providing the appropriate amount of background information was recognized as a challenge by the instructors too (see Sec.~\ref{sec:challenges}).

Students also discussed not liking when the experiment did not work as expected and they were not able to figure things out on their own. When asked what parts they did not enjoy, one student said ``every time it threw an error and I had to figure out what was happening.'' Another student discussed not enjoying ``whenever I got stuck,'' also saying ``I guess... if something doesn't work, that's not great. But it's still cool to... just to try to troubleshoot it. I guess if I was doing this a lot, then if Oqtant stopped working, that would be kind of annoying.'' The difficulty of troubleshooting was one of the drawbacks students noted about remote experiments (see Sec.~\ref{sec:comparison}), and that may be linked with students' lack of enjoyment when parts of the experiment or code did not work as expected. 

Overall, students enjoyed many aspects of working with Oqtant, but there were also some aspects they did not enjoy. Instructors are not able to control some of these aspects, such as the way Oqtant relies on job queues; however, instructors do have control over other aspects students mentioned. Instructors could consider providing students opportunities to explore the platform on their own, ensure students are provided sufficient background materials, and scaffold teaching how to troubleshoot issues the students may encounter.

\section{Conclusions}\label{sec:conclusions}

In this work, we investigated instructor and student views on a new publicly-available remote BEC experiment, as a way to begin studying possibilities to make quantum experiments accessible to students at institutions with any level of resources. Our primary goal was not evaluating student learning, so we cannot speak definitively to the efficacy of this approach; however, many instructors are excited about the possibility of remote quantum experiments, one instructor who incorporated Oqtant into their course perceived positive benefits for their students, and students' discussions of their experiences working with Oqtant are promising. This indicates that Oqtant and other similar types of remote experiments may have the potential to benefit students, although they will not replace hands-on experiences with apparatus.

Instructors foresee potential benefits to using Oqtant, but they also have concerns about implementing it in their courses. They are interested in incorporating Oqtant into different types of courses, and hope their students will use it to accomplish a range of learning outcomes including conceptual learning about quantum topics, student affect, and understanding how quantum theory manifests in experiments. Learning how to teach with a new experimental platform may also be a fun learning experience for instructors. However, using Oqtant in the classroom comes with challenges as well, including the need for reliable access to the experiment, the way neither instructors nor students are likely to be familiar with BECs, and the way it could replace an opportunity for a hands-on experience. While some of these challenges are inherent to any remote system, others may be able to be mitigated by support from various entities including outside educators, the company Infleqtion, and the instructors' institutions.

To help instructors consider ways to overcome these challenges, we presented a concrete example of one way Oqtant has been successfully incorporated into an upper-division quantum mechanics course and the benefits the instructor perceived for their students. Over a couple of class periods, the instructor helped their students work in groups on the two activities we developed and then assigned an open-ended project at the end of the semester. The instructor enjoyed the experience and is planning to implement it again when they next teach the same course if Oqtant is available. The instructor perceived a mix of responses from their students in terms of both enjoyment and comprehension, yet overall thought their students met their learning goals related to thinking critically about data and realizing experiments are not ideal.

The students' experiences also indicated that Oqtant may provide useful benefits. Although there was the possibility that students would not recognize Oqtant was a real experiment since they were only accessing it remotely, the students predominantly felt like they were working with a real experiment. This occurred both because they were receiving real experimental data (as evidenced partly from its variability) and because of the choices and actions the students were able to make. When comparing Oqtant with their prior experimental experiences, students noted both similarities and differences. These led to some student-perceived advantages, such as the way anyone could access Oqtant or the ease with which students could obtain a lot of real data. Students also noted drawbacks to remote experiments, although some of those may provide opportunities to teach students some skills that are authentic to physics research. Additionally, students enjoyed many aspects of working with Oqtant (e.g., exploring a platform with ultracold atoms), although there were some aspects of remote experiments (e.g., job queues and software installation) that they did not like. 

Comparing student and instructor views about the value they placed on Oqtant, we found both similarities and differences. One similarity is that many instructors wanted their students to have the opportunity to work with research-level experiments, and students did notice many ways that Oqtant is similar to apparatus in research labs. Additionally, some students really enjoyed working with Oqtant, which may contribute to the affective and motivational gains sought by instructors. On the other hand, the instructor of the upper-division course wanted students to experience the ``messiness of actual science,'' yet one of the aspects of working with Oqtant some students did not like was when things did not work as they expected. There may be a tension between some of instructors' goals for the students and students not necessarily recognizing the benefits they obtain when faced with unexpected results, an opportunity that may naturally arise from some types of experimental and computational activities \cite{phillips2021not, odden2023using}. Instructors could focus on helping students clearly articulate the problems they come across, as this is an important part of authentic scientific discovery \cite{phillips2017problematizing}.

This study points out potential ways both instructors and students think Oqtant could be useful in education and some of the perceived benefits, but further work needs to be done to rigorously evaluate if and how interacting with Oqtant can lead to these possible learning outcomes. In response to the many calls to train the upcoming quantum workforce, especially emphasizing the cultivation of experimental skills \cite{asfaw2022building, aiello2021achieving, fox2020preparing}, experiments like Oqtant may have the potential to teach some of these skills to a large number of students. Instructors and students in this study pointed out possible skills they could learn with Oqtant such as data analysis, coding, collaboration, dealing with unexpected results, and some types of troubleshooting. Some of these, such as data analysis, could come about from all of the less resource-intensive options discussed in Sec.~\ref{sec:bgAlternatives}; however, Oqtant could provide students the opportunity to practice their data collection skills and to analyze real experimental data they took themselves, which simulations and virtual screen experiments do not offer. Just like other remote quantum experiments, Oqtant may also provide students opportunities to learn how to design experimental procedures and troubleshoot experiments remotely, while ensuring that these opportunities are available for students at any institution. There is a dearth of studies demonstrating student development of experimental skills, so future work should investigate the efficacy of different proposed approaches, such as the use of remote cloud-accessible experiments, to supplement hands-on experiences and bring experimental experiences to students at a wider range of institutions.

If Oqtant or other similar remote experiments can provide a benefit to a wide range of students, the question then becomes: How do we ensure such an experiment remains free and publicly available? Even with the growing excitement in the educational community, Infleqtion recently announced an ``indefinite pause'' for Oqtant, and the future of the experiment is uncertain. What type of models of collaboration between academia and industry partners would allow for long-term sustainability of remote experiments while meeting the needs of students? Should educators continue to rely on industry to provide these remote experiments or are there resources available to ensure both the creation and sustainability of such experiments within the academic community \cite{barnes2024outcomes}? What other opportunities could exist for the establishment of free and publicly accessible quantum experiments that are tailored towards student learning?

This work is only the starting point of investigating new methods of providing access to quantum experiments to students everywhere. We hope both the educational materials we developed as well as this initial study of student and instructor experiences with Oqtant will serve as a model and help inform conversations about if such remote experiments would be beneficial to the educational community and how we can ensure their future existence. This would be just one step towards creating equitable learning opportunities to teach and inspire the next generation of quantum physicists.

\acknowledgements
We would like to thank the student and instructor participants in our study, especially the instructor who partnered with us to implement our activities in their course. We would also like to thank Alex Tingle, Noah Fitch, Anjul Loiacono, and the rest of the Oqtant team for partnering with us on this project. Lastly, we would like to thank the CU PER group for feedback and suggestions throughout this study. This work was supported by Infleqtion as well as NSF Grant PHY 2317149 and NSF QLCI Award OMA 2016244. The content of this work is solely the responsibility of the authors and does not necessarily represent the views of any of the funding sources.

\bibliography{OqtantResearch}

\begin{thebibliography}{93}%
\makeatletter
\providecommand \@ifxundefined [1]{%
 \@ifx{#1\undefined}
}%
\providecommand \@ifnum [1]{%
 \ifnum #1\expandafter \@firstoftwo
 \else \expandafter \@secondoftwo
 \fi
}%
\providecommand \@ifx [1]{%
 \ifx #1\expandafter \@firstoftwo
 \else \expandafter \@secondoftwo
 \fi
}%
\providecommand \natexlab [1]{#1}%
\providecommand \enquote  [1]{``#1''}%
\providecommand \bibnamefont  [1]{#1}%
\providecommand \bibfnamefont [1]{#1}%
\providecommand \citenamefont [1]{#1}%
\providecommand \href@noop [0]{\@secondoftwo}%
\providecommand \href [0]{\begingroup \@sanitize@url \@href}%
\providecommand \@href[1]{\@@startlink{#1}\@@href}%
\providecommand \@@href[1]{\endgroup#1\@@endlink}%
\providecommand \@sanitize@url [0]{\catcode `\\12\catcode `\$12\catcode
  `\&12\catcode `\#12\catcode `\^12\catcode `\_12\catcode `\%12\relax}%
\providecommand \@@startlink[1]{}%
\providecommand \@@endlink[0]{}%
\providecommand \url  [0]{\begingroup\@sanitize@url \@url }%
\providecommand \@url [1]{\endgroup\@href {#1}{\urlprefix }}%
\providecommand \urlprefix  [0]{URL }%
\providecommand \Eprint [0]{\href }%
\providecommand \doibase [0]{https://doi.org/}%
\providecommand \selectlanguage [0]{\@gobble}%
\providecommand \bibinfo  [0]{\@secondoftwo}%
\providecommand \bibfield  [0]{\@secondoftwo}%
\providecommand \translation [1]{[#1]}%
\providecommand \BibitemOpen [0]{}%
\providecommand \bibitemStop [0]{}%
\providecommand \bibitemNoStop [0]{.\EOS\space}%
\providecommand \EOS [0]{\spacefactor3000\relax}%
\providecommand \BibitemShut  [1]{\csname bibitem#1\endcsname}%
\let\auto@bib@innerbib\@empty
\bibitem [{nat()}]{nationalQuantumInitiativeAct}%
  \BibitemOpen
  \href@noop {} {}\bibinfo {note} {H.R. 6227 - National Quantum Initiative Act,
  115th Congress 164, 132 STAT. 5092, 2018,
  \url{https://www.congress.gov/115/plaws/publ368/PLAW-115publ368.pdf}}\BibitemShut
  {NoStop}%
\bibitem [{NQI(2022)}]{NQIsupplement}%
  \BibitemOpen
  \href
  {https://www.quantum.gov/wp-content/uploads/2022/02/QIST-Natl-Workforce-Plan.pdf}
  {\emph {\bibinfo {title} {Quantum Information Science and Technology
  Workforce Development National Strategic Plan}}},\ \bibinfo {type} {Tech.
  Rep.}\ (\bibinfo  {institution} {Subcommittee on Quantum Information Science,
  Committee on Science of the National Science \& Technology Council},\
  \bibinfo {year} {2022})\BibitemShut {NoStop}%
\bibitem [{\citenamefont {Riedel}\ \emph {et~al.}(2019)\citenamefont {Riedel},
  \citenamefont {Kovacs}, \citenamefont {Zoller}, \citenamefont {Mlynek},\ and\
  \citenamefont {Calarco}}]{riedel2019europe}%
  \BibitemOpen
  \bibfield  {author} {\bibinfo {author} {\bibfnamefont {M.}~\bibnamefont
  {Riedel}}, \bibinfo {author} {\bibfnamefont {M.}~\bibnamefont {Kovacs}},
  \bibinfo {author} {\bibfnamefont {P.}~\bibnamefont {Zoller}}, \bibinfo
  {author} {\bibfnamefont {J.}~\bibnamefont {Mlynek}},\ and\ \bibinfo {author}
  {\bibfnamefont {T.}~\bibnamefont {Calarco}},\ }\bibfield  {title} {\bibinfo
  {title} {Europe's quantum flagship initiative},\ }\href
  {https://iopscience.iop.org/article/10.1088/2058-9565/ab042d/meta} {\bibfield
   {journal} {\bibinfo  {journal} {Quant. Sci. Technol.}\ }\textbf {\bibinfo
  {volume} {4}},\ \bibinfo {pages} {020501} (\bibinfo {year}
  {2019})}\BibitemShut {NoStop}%
\bibitem [{\citenamefont {Perron}\ \emph {et~al.}(2021)\citenamefont {Perron},
  \citenamefont {DeLeone}, \citenamefont {Sharif}, \citenamefont {Carter},
  \citenamefont {Grossman}, \citenamefont {Passante},\ and\ \citenamefont
  {Sack}}]{perron2021quantum}%
  \BibitemOpen
  \bibfield  {author} {\bibinfo {author} {\bibfnamefont {J.~K.}\ \bibnamefont
  {Perron}}, \bibinfo {author} {\bibfnamefont {C.}~\bibnamefont {DeLeone}},
  \bibinfo {author} {\bibfnamefont {S.}~\bibnamefont {Sharif}}, \bibinfo
  {author} {\bibfnamefont {T.}~\bibnamefont {Carter}}, \bibinfo {author}
  {\bibfnamefont {J.~M.}\ \bibnamefont {Grossman}}, \bibinfo {author}
  {\bibfnamefont {G.}~\bibnamefont {Passante}},\ and\ \bibinfo {author}
  {\bibfnamefont {J.}~\bibnamefont {Sack}},\ }\bibfield  {title} {\bibinfo
  {title} {Quantum undergraduate education and scientific training},\ }\href
  {https://arxiv.org/abs/2109.13850} {\bibfield  {journal} {\bibinfo  {journal}
  {arXiv:2109.13850}\ } (\bibinfo {year} {2021})}\BibitemShut {NoStop}%
\bibitem [{\citenamefont {Meyer}\ \emph {et~al.}(2022)\citenamefont {Meyer},
  \citenamefont {Passante}, \citenamefont {Pollock},\ and\ \citenamefont
  {Wilcox}}]{meyer2022todays}%
  \BibitemOpen
  \bibfield  {author} {\bibinfo {author} {\bibfnamefont {J.~C.}\ \bibnamefont
  {Meyer}}, \bibinfo {author} {\bibfnamefont {G.}~\bibnamefont {Passante}},
  \bibinfo {author} {\bibfnamefont {S.~J.}\ \bibnamefont {Pollock}},\ and\
  \bibinfo {author} {\bibfnamefont {B.~R.}\ \bibnamefont {Wilcox}},\ }\bibfield
   {title} {\bibinfo {title} {Today's interdisciplinary quantum information
  classroom: Themes from a survey of quantum information science instructors},\
  }\href {https://doi.org/10.1103/PhysRevPhysEducRes.18.010150} {\bibfield
  {journal} {\bibinfo  {journal} {Phys. Rev. Phys. Educ. Res.}\ }\textbf
  {\bibinfo {volume} {18}},\ \bibinfo {pages} {010150} (\bibinfo {year}
  {2022})}\BibitemShut {NoStop}%
\bibitem [{\citenamefont {Kaur}\ and\ \citenamefont
  {Venegas-Gomez}(2022)}]{kaur2022defining}%
  \BibitemOpen
  \bibfield  {author} {\bibinfo {author} {\bibfnamefont {M.}~\bibnamefont
  {Kaur}}\ and\ \bibinfo {author} {\bibfnamefont {A.}~\bibnamefont
  {Venegas-Gomez}},\ }\bibfield  {title} {\bibinfo {title} {Defining the
  quantum workforce landscape: {A} review of global quantum education
  initiatives},\ }\href {https://doi.org/10.1117/1.OE.61.8.081806} {\bibfield
  {journal} {\bibinfo  {journal} {Opt. Eng.}\ }\textbf {\bibinfo {volume}
  {61}},\ \bibinfo {pages} {081806} (\bibinfo {year} {2022})}\BibitemShut
  {NoStop}%
\bibitem [{\citenamefont {Hasanovic}\ \emph {et~al.}(2022)\citenamefont
  {Hasanovic}, \citenamefont {Panayiotou}, \citenamefont {Silberman},
  \citenamefont {Stimers},\ and\ \citenamefont
  {Merzbacher}}]{hasanovic2022quantum}%
  \BibitemOpen
  \bibfield  {author} {\bibinfo {author} {\bibfnamefont {M.}~\bibnamefont
  {Hasanovic}}, \bibinfo {author} {\bibfnamefont {C.}~\bibnamefont
  {Panayiotou}}, \bibinfo {author} {\bibfnamefont {D.}~\bibnamefont
  {Silberman}}, \bibinfo {author} {\bibfnamefont {P.}~\bibnamefont {Stimers}},\
  and\ \bibinfo {author} {\bibfnamefont {C.}~\bibnamefont {Merzbacher}},\
  }\bibfield  {title} {\bibinfo {title} {Quantum technician skills and
  competencies for the emerging {Q}uantum 2.0 industry},\ }\href
  {https://doi.org/10.1117/1.OE.61.8.081803} {\bibfield  {journal} {\bibinfo
  {journal} {Opt. Eng.}\ }\textbf {\bibinfo {volume} {61}},\ \bibinfo {pages}
  {081803} (\bibinfo {year} {2022})}\BibitemShut {NoStop}%
\bibitem [{\citenamefont {Stadermann}\ \emph {et~al.}(2019)\citenamefont
  {Stadermann}, \citenamefont {van~den Berg},\ and\ \citenamefont
  {Goedhart}}]{stadermann2019analysis}%
  \BibitemOpen
  \bibfield  {author} {\bibinfo {author} {\bibfnamefont {H.~K.~E.}\
  \bibnamefont {Stadermann}}, \bibinfo {author} {\bibfnamefont
  {E.}~\bibnamefont {van~den Berg}},\ and\ \bibinfo {author} {\bibfnamefont
  {M.~J.}\ \bibnamefont {Goedhart}},\ }\bibfield  {title} {\bibinfo {title}
  {Analysis of secondary school quantum physics curricula of 15 different
  countries: Different perspectives on a challenging topic},\ }\href
  {https://doi.org/10.1103/PhysRevPhysEducRes.15.010130} {\bibfield  {journal}
  {\bibinfo  {journal} {Phys. Rev. Phys. Educ. Res.}\ }\textbf {\bibinfo
  {volume} {15}},\ \bibinfo {pages} {010130} (\bibinfo {year}
  {2019})}\BibitemShut {NoStop}%
\bibitem [{\citenamefont {Seskir}\ \emph {et~al.}(2022)\citenamefont {Seskir},
  \citenamefont {Migda{\l}}, \citenamefont {Weidner}, \citenamefont {Anupam},
  \citenamefont {Case}, \citenamefont {Davis}, \citenamefont {Decaroli},
  \citenamefont {Ercan}, \citenamefont {Foti}, \citenamefont {Gora} \emph
  {et~al.}}]{seskir2022quantum}%
  \BibitemOpen
  \bibfield  {author} {\bibinfo {author} {\bibfnamefont {Z.~C.}\ \bibnamefont
  {Seskir}}, \bibinfo {author} {\bibfnamefont {P.}~\bibnamefont {Migda{\l}}},
  \bibinfo {author} {\bibfnamefont {C.}~\bibnamefont {Weidner}}, \bibinfo
  {author} {\bibfnamefont {A.}~\bibnamefont {Anupam}}, \bibinfo {author}
  {\bibfnamefont {N.}~\bibnamefont {Case}}, \bibinfo {author} {\bibfnamefont
  {N.}~\bibnamefont {Davis}}, \bibinfo {author} {\bibfnamefont
  {C.}~\bibnamefont {Decaroli}}, \bibinfo {author} {\bibfnamefont
  {{\.I}.}~\bibnamefont {Ercan}}, \bibinfo {author} {\bibfnamefont
  {C.}~\bibnamefont {Foti}}, \bibinfo {author} {\bibfnamefont {P.}~\bibnamefont
  {Gora}}, \emph {et~al.},\ }\bibfield  {title} {\bibinfo {title} {Quantum
  games and interactive tools for quantum technologies outreach and
  education},\ }\href {https://doi.org/10.1117/1.OE.61.8.081809} {\bibfield
  {journal} {\bibinfo  {journal} {Optical Engineering}\ }\textbf {\bibinfo
  {volume} {61}},\ \bibinfo {pages} {081809} (\bibinfo {year}
  {2022})}\BibitemShut {NoStop}%
\bibitem [{\citenamefont {Asfaw}\ \emph {et~al.}(2022)\citenamefont {Asfaw},
  \citenamefont {Blais}, \citenamefont {Brown}, \citenamefont {Candelaria},
  \citenamefont {Cantwell}, \citenamefont {Carr}, \citenamefont {Combes},
  \citenamefont {Debroy}, \citenamefont {Donohue}, \citenamefont {Economou}
  \emph {et~al.}}]{asfaw2022building}%
  \BibitemOpen
  \bibfield  {author} {\bibinfo {author} {\bibfnamefont {A.}~\bibnamefont
  {Asfaw}}, \bibinfo {author} {\bibfnamefont {A.}~\bibnamefont {Blais}},
  \bibinfo {author} {\bibfnamefont {K.~R.}\ \bibnamefont {Brown}}, \bibinfo
  {author} {\bibfnamefont {J.}~\bibnamefont {Candelaria}}, \bibinfo {author}
  {\bibfnamefont {C.}~\bibnamefont {Cantwell}}, \bibinfo {author}
  {\bibfnamefont {L.~D.}\ \bibnamefont {Carr}}, \bibinfo {author}
  {\bibfnamefont {J.}~\bibnamefont {Combes}}, \bibinfo {author} {\bibfnamefont
  {D.~M.}\ \bibnamefont {Debroy}}, \bibinfo {author} {\bibfnamefont {J.~M.}\
  \bibnamefont {Donohue}}, \bibinfo {author} {\bibfnamefont {S.~E.}\
  \bibnamefont {Economou}}, \emph {et~al.},\ }\bibfield  {title} {\bibinfo
  {title} {Building a quantum engineering undergraduate program},\ }\href
  {https://ieeexplore.ieee.org/document/9705217} {\bibfield  {journal}
  {\bibinfo  {journal} {IEEE Trans. Educ.}\ }\textbf {\bibinfo {volume} {65}}
  (\bibinfo {year} {2022})}\BibitemShut {NoStop}%
\bibitem [{\citenamefont {Aiello}\ \emph {et~al.}(2021)\citenamefont {Aiello},
  \citenamefont {Awschalom}, \citenamefont {Bernien}, \citenamefont {Brower},
  \citenamefont {Brown}, \citenamefont {Brun}, \citenamefont {Caram},
  \citenamefont {Chitambar}, \citenamefont {Di~Felice}, \citenamefont {Edmonds}
  \emph {et~al.}}]{aiello2021achieving}%
  \BibitemOpen
  \bibfield  {author} {\bibinfo {author} {\bibfnamefont {C.~D.}\ \bibnamefont
  {Aiello}}, \bibinfo {author} {\bibfnamefont {D.}~\bibnamefont {Awschalom}},
  \bibinfo {author} {\bibfnamefont {H.}~\bibnamefont {Bernien}}, \bibinfo
  {author} {\bibfnamefont {T.}~\bibnamefont {Brower}}, \bibinfo {author}
  {\bibfnamefont {K.~R.}\ \bibnamefont {Brown}}, \bibinfo {author}
  {\bibfnamefont {T.~A.}\ \bibnamefont {Brun}}, \bibinfo {author}
  {\bibfnamefont {J.~R.}\ \bibnamefont {Caram}}, \bibinfo {author}
  {\bibfnamefont {E.}~\bibnamefont {Chitambar}}, \bibinfo {author}
  {\bibfnamefont {R.}~\bibnamefont {Di~Felice}}, \bibinfo {author}
  {\bibfnamefont {K.~M.}\ \bibnamefont {Edmonds}}, \emph {et~al.},\ }\bibfield
  {title} {\bibinfo {title} {Achieving a quantum smart workforce},\ }\href
  {https://iopscience.iop.org/article/10.1088/2058-9565/abfa64/meta} {\bibfield
   {journal} {\bibinfo  {journal} {Quant. Sci. Technol.}\ }\textbf {\bibinfo
  {volume} {6}},\ \bibinfo {pages} {030501} (\bibinfo {year}
  {2021})}\BibitemShut {NoStop}%
\bibitem [{\citenamefont {Fox}\ \emph {et~al.}(2020)\citenamefont {Fox},
  \citenamefont {Zwickl},\ and\ \citenamefont
  {Lewandowski}}]{fox2020preparing}%
  \BibitemOpen
  \bibfield  {author} {\bibinfo {author} {\bibfnamefont {M.~F.~J.}\
  \bibnamefont {Fox}}, \bibinfo {author} {\bibfnamefont {B.~M.}\ \bibnamefont
  {Zwickl}},\ and\ \bibinfo {author} {\bibfnamefont {H.~J.}\ \bibnamefont
  {Lewandowski}},\ }\bibfield  {title} {\bibinfo {title} {Preparing for the
  quantum revolution: What is the role of higher education?},\ }\href
  {https://doi.org/10.1103/PhysRevPhysEducRes.16.020131} {\bibfield  {journal}
  {\bibinfo  {journal} {Phys. Rev. Phys. Educ. Res.}\ }\textbf {\bibinfo
  {volume} {16}},\ \bibinfo {pages} {020131} (\bibinfo {year}
  {2020})}\BibitemShut {NoStop}%
\bibitem [{\citenamefont {Greinert}\ \emph {et~al.}(2024)\citenamefont
  {Greinert}, \citenamefont {Ubben}, \citenamefont {Dogan}, \citenamefont
  {Hilfert-R{\"u}ppell},\ and\ \citenamefont
  {M{\"u}ller}}]{greinert2024advancing}%
  \BibitemOpen
  \bibfield  {author} {\bibinfo {author} {\bibfnamefont {F.}~\bibnamefont
  {Greinert}}, \bibinfo {author} {\bibfnamefont {M.~S.}\ \bibnamefont {Ubben}},
  \bibinfo {author} {\bibfnamefont {I.~N.}\ \bibnamefont {Dogan}}, \bibinfo
  {author} {\bibfnamefont {D.}~\bibnamefont {Hilfert-R{\"u}ppell}},\ and\
  \bibinfo {author} {\bibfnamefont {R.}~\bibnamefont {M{\"u}ller}},\ }\bibfield
   {title} {\bibinfo {title} {Advancing quantum technology workforce: industry
  insights into qualification and training needs},\ }\href
  {https://doi.org/10.48550/arXiv.2407.21598} {\bibfield  {journal} {\bibinfo
  {journal} {arXiv:2407.21598}\ } (\bibinfo {year} {2024})}\BibitemShut
  {NoStop}%
\bibitem [{\citenamefont {Hughes}\ \emph {et~al.}(2022)\citenamefont {Hughes},
  \citenamefont {Finke}, \citenamefont {German}, \citenamefont {Merzbacher},
  \citenamefont {Vora},\ and\ \citenamefont
  {Lewandowski}}]{hughes2022assessing}%
  \BibitemOpen
  \bibfield  {author} {\bibinfo {author} {\bibfnamefont {C.}~\bibnamefont
  {Hughes}}, \bibinfo {author} {\bibfnamefont {D.}~\bibnamefont {Finke}},
  \bibinfo {author} {\bibfnamefont {D.-A.}\ \bibnamefont {German}}, \bibinfo
  {author} {\bibfnamefont {C.}~\bibnamefont {Merzbacher}}, \bibinfo {author}
  {\bibfnamefont {P.~M.}\ \bibnamefont {Vora}},\ and\ \bibinfo {author}
  {\bibfnamefont {H.}~\bibnamefont {Lewandowski}},\ }\bibfield  {title}
  {\bibinfo {title} {Assessing the needs of the quantum industry},\ }\href
  {https://ieeexplore.ieee.org/abstract/document/9733176} {\bibfield  {journal}
  {\bibinfo  {journal} {IEEE Trans. Educ.}\ } (\bibinfo {year}
  {2022})}\BibitemShut {NoStop}%
\bibitem [{\citenamefont {Greinert}\ \emph {et~al.}(2023)\citenamefont
  {Greinert}, \citenamefont {M\"uller}, \citenamefont {Bitzenbauer},
  \citenamefont {Ubben},\ and\ \citenamefont {Weber}}]{greinert2023future}%
  \BibitemOpen
  \bibfield  {author} {\bibinfo {author} {\bibfnamefont {F.}~\bibnamefont
  {Greinert}}, \bibinfo {author} {\bibfnamefont {R.}~\bibnamefont {M\"uller}},
  \bibinfo {author} {\bibfnamefont {P.}~\bibnamefont {Bitzenbauer}}, \bibinfo
  {author} {\bibfnamefont {M.~S.}\ \bibnamefont {Ubben}},\ and\ \bibinfo
  {author} {\bibfnamefont {K.-A.}\ \bibnamefont {Weber}},\ }\bibfield  {title}
  {\bibinfo {title} {Future quantum workforce: Competences, requirements, and
  forecasts},\ }\href {https://doi.org/10.1103/PhysRevPhysEducRes.19.010137}
  {\bibfield  {journal} {\bibinfo  {journal} {Phys. Rev. Phys. Educ. Res.}\
  }\textbf {\bibinfo {volume} {19}},\ \bibinfo {pages} {010137} (\bibinfo
  {year} {2023})}\BibitemShut {NoStop}%
\bibitem [{\citenamefont {Kozminski}\ \emph {et~al.}(2014)\citenamefont
  {Kozminski}, \citenamefont {Lewandowski}, \citenamefont {Beverly},
  \citenamefont {Lindaas}, \citenamefont {Deardorff}, \citenamefont {Reagan},
  \citenamefont {Dietz}, \citenamefont {Tagg}, \citenamefont {Eblen-Zayas},
  \citenamefont {Williams} \emph {et~al.}}]{kozminski2014aapt}%
  \BibitemOpen
  \bibfield  {author} {\bibinfo {author} {\bibfnamefont {J.}~\bibnamefont
  {Kozminski}}, \bibinfo {author} {\bibfnamefont {H.}~\bibnamefont
  {Lewandowski}}, \bibinfo {author} {\bibfnamefont {N.}~\bibnamefont
  {Beverly}}, \bibinfo {author} {\bibfnamefont {S.}~\bibnamefont {Lindaas}},
  \bibinfo {author} {\bibfnamefont {D.}~\bibnamefont {Deardorff}}, \bibinfo
  {author} {\bibfnamefont {A.}~\bibnamefont {Reagan}}, \bibinfo {author}
  {\bibfnamefont {R.}~\bibnamefont {Dietz}}, \bibinfo {author} {\bibfnamefont
  {R.}~\bibnamefont {Tagg}}, \bibinfo {author} {\bibfnamefont {M.}~\bibnamefont
  {Eblen-Zayas}}, \bibinfo {author} {\bibfnamefont {J.}~\bibnamefont
  {Williams}}, \emph {et~al.},\ }\bibfield  {title} {\bibinfo {title} {{AAPT}
  recommendations for the undergraduate physics laboratory curriculum},\ }\href
  {https://www.aapt.org/resources/upload/labguidlinesdocument_ebendorsed_nov10.pdf}
  {\bibfield  {journal} {\bibinfo  {journal} {American Association of Physics
  Teachers}\ } (\bibinfo {year} {2014})}\BibitemShut {NoStop}%
\bibitem [{\citenamefont {{Joint Task Force on Undergraduate Physics
  Programs}}(2016)}]{joint2016phys21}%
  \BibitemOpen
  \bibfield  {author} {\bibinfo {author} {\bibnamefont {{Joint Task Force on
  Undergraduate Physics Programs}}},\ }\href
  {https://www.compadre.org/jtupp/report.cfm} {\bibinfo {title} {Phys21:
  {P}reparing physics student for 21st-century careers}} (\bibinfo {year}
  {2016})\BibitemShut {NoStop}%
\bibitem [{\citenamefont {Auchincloss}\ \emph {et~al.}(2014)\citenamefont
  {Auchincloss}, \citenamefont {Laursen}, \citenamefont {Branchaw},
  \citenamefont {Eagan}, \citenamefont {Graham}, \citenamefont {Hanauer},
  \citenamefont {Lawrie}, \citenamefont {McLinn}, \citenamefont {Pelaez},
  \citenamefont {Rowland} \emph {et~al.}}]{auchincloss2014assessment}%
  \BibitemOpen
  \bibfield  {author} {\bibinfo {author} {\bibfnamefont {L.~C.}\ \bibnamefont
  {Auchincloss}}, \bibinfo {author} {\bibfnamefont {S.~L.}\ \bibnamefont
  {Laursen}}, \bibinfo {author} {\bibfnamefont {J.~L.}\ \bibnamefont
  {Branchaw}}, \bibinfo {author} {\bibfnamefont {K.}~\bibnamefont {Eagan}},
  \bibinfo {author} {\bibfnamefont {M.}~\bibnamefont {Graham}}, \bibinfo
  {author} {\bibfnamefont {D.~I.}\ \bibnamefont {Hanauer}}, \bibinfo {author}
  {\bibfnamefont {G.}~\bibnamefont {Lawrie}}, \bibinfo {author} {\bibfnamefont
  {C.~M.}\ \bibnamefont {McLinn}}, \bibinfo {author} {\bibfnamefont
  {N.}~\bibnamefont {Pelaez}}, \bibinfo {author} {\bibfnamefont
  {S.}~\bibnamefont {Rowland}}, \emph {et~al.},\ }\href
  {https://doi.org/10.1187/cbe.14-01-0004} {\bibinfo {title} {Assessment of
  course-based undergraduate research experiences: a meeting report}} (\bibinfo
  {year} {2014})\BibitemShut {NoStop}%
\bibitem [{\citenamefont {Buchanan}\ and\ \citenamefont
  {Fisher}(2022)}]{buchanan2022current}%
  \BibitemOpen
  \bibfield  {author} {\bibinfo {author} {\bibfnamefont {A.~J.}\ \bibnamefont
  {Buchanan}}\ and\ \bibinfo {author} {\bibfnamefont {G.~R.}\ \bibnamefont
  {Fisher}},\ }\bibfield  {title} {\bibinfo {title} {Current status and
  implementation of science practices in course-based undergraduate research
  experiences ({CURE}s): {A} systematic literature review},\ }\href
  {10.1187/cbe.22-04-0069} {\bibfield  {journal} {\bibinfo  {journal}
  {CBE—Life Sciences Education}\ }\textbf {\bibinfo {volume} {21}},\ \bibinfo
  {pages} {ar83} (\bibinfo {year} {2022})}\BibitemShut {NoStop}%
\bibitem [{\citenamefont {Merritt}\ and\ \citenamefont
  {Lewandowski}(2024)}]{merritt2024physics}%
  \BibitemOpen
  \bibfield  {author} {\bibinfo {author} {\bibfnamefont {R.~L.}\ \bibnamefont
  {Merritt}}\ and\ \bibinfo {author} {\bibfnamefont {H.~J.}\ \bibnamefont
  {Lewandowski}},\ }\bibfield  {title} {\bibinfo {title} {Physics instructor
  views on course-based undergraduate research experiences (cures)},\ }in\
  \href {https://doi.org/10.1119/perc.2024.pr.Merritt.} {\emph {\bibinfo
  {booktitle} {2024 PERC Proceedings}}},\ \bibinfo {editor} {edited by\
  \bibinfo {editor} {\bibfnamefont {Q.~X.}\ \bibnamefont {Ryan}}, \bibinfo
  {editor} {\bibfnamefont {A.}~\bibnamefont {Pawl}},\ and\ \bibinfo {editor}
  {\bibfnamefont {J.~P.}\ \bibnamefont {Zwolak}}}\ (\bibinfo {year}
  {2024})\BibitemShut {NoStop}%
\bibitem [{\citenamefont {Smith}\ \emph {et~al.}(2020)\citenamefont {Smith},
  \citenamefont {Stein}, \citenamefont {Walsh},\ and\ \citenamefont
  {Holmes}}]{smith2020direct}%
  \BibitemOpen
  \bibfield  {author} {\bibinfo {author} {\bibfnamefont {E.~M.}\ \bibnamefont
  {Smith}}, \bibinfo {author} {\bibfnamefont {M.~M.}\ \bibnamefont {Stein}},
  \bibinfo {author} {\bibfnamefont {C.}~\bibnamefont {Walsh}},\ and\ \bibinfo
  {author} {\bibfnamefont {N.~G.}\ \bibnamefont {Holmes}},\ }\bibfield  {title}
  {\bibinfo {title} {Direct measurement of the impact of teaching
  experimentation in physics labs},\ }\href
  {https://doi.org/10.1103/PhysRevX.10.011029} {\bibfield  {journal} {\bibinfo
  {journal} {Phys. Rev. X}\ }\textbf {\bibinfo {volume} {10}},\ \bibinfo
  {pages} {011029} (\bibinfo {year} {2020})}\BibitemShut {NoStop}%
\bibitem [{\citenamefont {Dutson}\ \emph {et~al.}(1997)\citenamefont {Dutson},
  \citenamefont {Todd}, \citenamefont {Magleby},\ and\ \citenamefont
  {Sorensen}}]{dutson1997review}%
  \BibitemOpen
  \bibfield  {author} {\bibinfo {author} {\bibfnamefont {A.~J.}\ \bibnamefont
  {Dutson}}, \bibinfo {author} {\bibfnamefont {R.~H.}\ \bibnamefont {Todd}},
  \bibinfo {author} {\bibfnamefont {S.~P.}\ \bibnamefont {Magleby}},\ and\
  \bibinfo {author} {\bibfnamefont {C.~D.}\ \bibnamefont {Sorensen}},\
  }\bibfield  {title} {\bibinfo {title} {A review of literature on teaching
  engineering design through project-oriented capstone courses},\ }\href
  {https://doi.org/10.1002/j.2168-9830.1997.tb00260.x} {\bibfield  {journal}
  {\bibinfo  {journal} {Journal of engineering education}\ }\textbf {\bibinfo
  {volume} {86}},\ \bibinfo {pages} {17} (\bibinfo {year} {1997})}\BibitemShut
  {NoStop}%
\bibitem [{\citenamefont {Holmes}\ and\ \citenamefont
  {Wieman}(2016)}]{holmes2016examining}%
  \BibitemOpen
  \bibfield  {author} {\bibinfo {author} {\bibfnamefont {N.~G.}\ \bibnamefont
  {Holmes}}\ and\ \bibinfo {author} {\bibfnamefont {C.~E.}\ \bibnamefont
  {Wieman}},\ }\bibfield  {title} {\bibinfo {title} {Examining and contrasting
  the cognitive activities engaged in undergraduate research experiences and
  lab courses},\ }\href {https://doi.org/10.1103/PhysRevPhysEducRes.12.020103}
  {\bibfield  {journal} {\bibinfo  {journal} {Phys. Rev. Phys. Educ. Res.}\
  }\textbf {\bibinfo {volume} {12}},\ \bibinfo {pages} {020103} (\bibinfo
  {year} {2016})}\BibitemShut {NoStop}%
\bibitem [{\citenamefont {Oliver}\ \emph {et~al.}(2023)\citenamefont {Oliver},
  \citenamefont {Werth},\ and\ \citenamefont
  {Lewandowski}}]{oliver2023student}%
  \BibitemOpen
  \bibfield  {author} {\bibinfo {author} {\bibfnamefont {K.~A.}\ \bibnamefont
  {Oliver}}, \bibinfo {author} {\bibfnamefont {A.}~\bibnamefont {Werth}},\ and\
  \bibinfo {author} {\bibfnamefont {H.~J.}\ \bibnamefont {Lewandowski}},\
  }\bibfield  {title} {\bibinfo {title} {Student experiences with authentic
  research in a remote, introductory course-based undergraduate research
  experience in physics},\ }\href
  {https://doi.org/10.1103/PhysRevPhysEducRes.19.010124} {\bibfield  {journal}
  {\bibinfo  {journal} {Phys. Rev. Phys. Educ. Res.}\ }\textbf {\bibinfo
  {volume} {19}},\ \bibinfo {pages} {010124} (\bibinfo {year}
  {2023})}\BibitemShut {NoStop}%
\bibitem [{\citenamefont {Meyer}\ \emph {et~al.}(2024)\citenamefont {Meyer},
  \citenamefont {Passante},\ and\ \citenamefont
  {Wilcox}}]{meyer2024disparities}%
  \BibitemOpen
  \bibfield  {author} {\bibinfo {author} {\bibfnamefont {J.~C.}\ \bibnamefont
  {Meyer}}, \bibinfo {author} {\bibfnamefont {G.}~\bibnamefont {Passante}},\
  and\ \bibinfo {author} {\bibfnamefont {B.}~\bibnamefont {Wilcox}},\
  }\bibfield  {title} {\bibinfo {title} {Disparities in access to u.s. quantum
  information education},\ }\href
  {https://doi.org/10.1103/PhysRevPhysEducRes.20.010131} {\bibfield  {journal}
  {\bibinfo  {journal} {Phys. Rev. Phys. Educ. Res.}\ }\textbf {\bibinfo
  {volume} {20}},\ \bibinfo {pages} {010131} (\bibinfo {year}
  {2024})}\BibitemShut {NoStop}%
\bibitem [{\citenamefont {Rosenberg}\ \emph {et~al.}(2024)\citenamefont
  {Rosenberg}, \citenamefont {Holincheck},\ and\ \citenamefont
  {Colandene}}]{rosenberg2024science}%
  \BibitemOpen
  \bibfield  {author} {\bibinfo {author} {\bibfnamefont {J.~L.}\ \bibnamefont
  {Rosenberg}}, \bibinfo {author} {\bibfnamefont {N.}~\bibnamefont
  {Holincheck}},\ and\ \bibinfo {author} {\bibfnamefont {M.}~\bibnamefont
  {Colandene}},\ }\bibfield  {title} {\bibinfo {title} {Science, technology,
  engineering, and mathematics undergraduates' knowledge and interest in
  quantum careers: Barriers and opportunities to building a diverse quantum
  workforce},\ }\href {https://doi.org/10.1103/PhysRevPhysEducRes.20.010138}
  {\bibfield  {journal} {\bibinfo  {journal} {Phys. Rev. Phys. Educ. Res.}\
  }\textbf {\bibinfo {volume} {20}},\ \bibinfo {pages} {010138} (\bibinfo
  {year} {2024})}\BibitemShut {NoStop}%
\bibitem [{\citenamefont {Bennett}\ \emph {et~al.}(2024)\citenamefont
  {Bennett}, \citenamefont {Arrow}, \citenamefont {Novack},\ and\ \citenamefont
  {Finkelstein}}]{bennett2024investigating}%
  \BibitemOpen
  \bibfield  {author} {\bibinfo {author} {\bibfnamefont {M.~B.}\ \bibnamefont
  {Bennett}}, \bibinfo {author} {\bibfnamefont {J.~{\'E}.}\ \bibnamefont
  {Arrow}}, \bibinfo {author} {\bibfnamefont {S.}~\bibnamefont {Novack}},\ and\
  \bibinfo {author} {\bibfnamefont {N.~D.}\ \bibnamefont {Finkelstein}},\
  }\bibfield  {title} {\bibinfo {title} {Investigating student participation in
  quantum workforce initiatives},\ }\href
  {https://doi.org/10.48550/arXiv.2407.14698} {\bibfield  {journal} {\bibinfo
  {journal} {arXiv:2407.14698}\ } (\bibinfo {year} {2024})}\BibitemShut
  {NoStop}%
\bibitem [{\citenamefont {Anderson}\ \emph {et~al.}(1995)\citenamefont
  {Anderson}, \citenamefont {Ensher}, \citenamefont {Matthews}, \citenamefont
  {Wieman},\ and\ \citenamefont {Cornell}}]{anderson1995observation}%
  \BibitemOpen
  \bibfield  {author} {\bibinfo {author} {\bibfnamefont {M.~H.}\ \bibnamefont
  {Anderson}}, \bibinfo {author} {\bibfnamefont {J.~R.}\ \bibnamefont
  {Ensher}}, \bibinfo {author} {\bibfnamefont {M.~R.}\ \bibnamefont
  {Matthews}}, \bibinfo {author} {\bibfnamefont {C.~E.}\ \bibnamefont
  {Wieman}},\ and\ \bibinfo {author} {\bibfnamefont {E.~A.}\ \bibnamefont
  {Cornell}},\ }\bibfield  {title} {\bibinfo {title} {Observation of
  {B}ose-{E}instein condensation in a dilute atomic vapor},\ }\href
  {https://doi.org/10.1126/science.269.5221.198} {\bibfield  {journal}
  {\bibinfo  {journal} {Science}\ }\textbf {\bibinfo {volume} {269}},\ \bibinfo
  {pages} {198} (\bibinfo {year} {1995})}\BibitemShut {NoStop}%
\bibitem [{oqt({\natexlab{a}})}]{oqtantWebsite}%
  \BibitemOpen
  \href@noop {} {} ({\natexlab{a}}),\ \bibinfo {note} {{O}qtant,
  \url{https://oqtant.infleqtion.com/}}\BibitemShut {NoStop}%
\bibitem [{\citenamefont {Tingle}\ \emph {et~al.}(2024)\citenamefont {Tingle},
  \citenamefont {Loiacono}, \citenamefont {Colussi},\ and\ \citenamefont
  {Fitch}}]{tingle2024oqtant}%
  \BibitemOpen
  \bibfield  {author} {\bibinfo {author} {\bibfnamefont {A.~E.}\ \bibnamefont
  {Tingle}}, \bibinfo {author} {\bibfnamefont {A.}~\bibnamefont {Loiacono}},
  \bibinfo {author} {\bibfnamefont {V.~E.}\ \bibnamefont {Colussi}},\ and\
  \bibinfo {author} {\bibfnamefont {N.}~\bibnamefont {Fitch}},\ }\bibfield
  {title} {\bibinfo {title} {Oqtant: democratizing quantum research and
  education},\ }in\ \href {https://doi.org/10.1117/12.3003525} {\emph {\bibinfo
  {booktitle} {Quantum Computing, Communication, and Simulation IV}}},\ Vol.\
  \bibinfo {volume} {12911}\ (\bibinfo {organization} {SPIE},\ \bibinfo {year}
  {2024})\ pp.\ \bibinfo {pages} {345--352}\BibitemShut {NoStop}%
\bibitem [{\citenamefont {Casta{\~n}o}\ \emph {et~al.}(2024)\citenamefont
  {Casta{\~n}o}, \citenamefont {L{\'o}pez}, \citenamefont {Jaramillo},
  \citenamefont {Navarro},\ and\ \citenamefont
  {Osorio}}]{castano2024deploying}%
  \BibitemOpen
  \bibfield  {author} {\bibinfo {author} {\bibfnamefont {F.~A.}\ \bibnamefont
  {Casta{\~n}o}}, \bibinfo {author} {\bibfnamefont {E.}~\bibnamefont
  {L{\'o}pez}}, \bibinfo {author} {\bibfnamefont {J.~A.}\ \bibnamefont
  {Jaramillo}}, \bibinfo {author} {\bibfnamefont {V.}~\bibnamefont {Navarro}},\
  and\ \bibinfo {author} {\bibfnamefont {J.}~\bibnamefont {Osorio}},\
  }\bibfield  {title} {\bibinfo {title} {Deploying an {IoT}-based remote
  physics lab platform to enhance experimental physics education in remote
  regions},\ }\href {https://doi.org/10.1088/1361-6552/ad7a47} {\bibfield
  {journal} {\bibinfo  {journal} {Physics Education}\ }\textbf {\bibinfo
  {volume} {59}},\ \bibinfo {pages} {065017} (\bibinfo {year}
  {2024})}\BibitemShut {NoStop}%
\bibitem [{rem()}]{remoteGlowDischargeExpt}%
  \BibitemOpen
  \href@noop {} {}\bibinfo {note} {Princeton Plasma Physics Lab, Remote Glow
  Discharge Experiment.
  \url{https://www.pppl.gov/remote-glow-discharge-experiment-rgdx}.}\BibitemShut
  {Stop}%
\bibitem [{IBM()}]{IBMquantumComputing}%
  \BibitemOpen
  \href@noop {} {}\bibinfo {note} {IBM Quantum Computing,
  \url{https://www.ibm.com/quantum}}\BibitemShut {NoStop}%
\bibitem [{\citenamefont {Kohnle}\ and\ \citenamefont
  {Rizzoli}(2017)}]{kohnle2017interactive}%
  \BibitemOpen
  \bibfield  {author} {\bibinfo {author} {\bibfnamefont {A.}~\bibnamefont
  {Kohnle}}\ and\ \bibinfo {author} {\bibfnamefont {A.}~\bibnamefont
  {Rizzoli}},\ }\bibfield  {title} {\bibinfo {title} {Interactive simulations
  for quantum key distribution},\ }\href
  {https://doi.org/10.1088/1361-6404/aa62c8} {\bibfield  {journal} {\bibinfo
  {journal} {European Journal of Physics}\ }\textbf {\bibinfo {volume} {38}},\
  \bibinfo {pages} {035403} (\bibinfo {year} {2017})}\BibitemShut {NoStop}%
\bibitem [{\citenamefont {Hu}\ \emph {et~al.}(2024)\citenamefont {Hu},
  \citenamefont {Li},\ and\ \citenamefont {Singh}}]{hu2024investigating}%
  \BibitemOpen
  \bibfield  {author} {\bibinfo {author} {\bibfnamefont {P.}~\bibnamefont
  {Hu}}, \bibinfo {author} {\bibfnamefont {Y.}~\bibnamefont {Li}},\ and\
  \bibinfo {author} {\bibfnamefont {C.}~\bibnamefont {Singh}},\ }\bibfield
  {title} {\bibinfo {title} {Investigating and improving student understanding
  of the basics of quantum computing},\ }\href
  {https://doi.org/10.1103/PhysRevPhysEducRes.20.020108} {\bibfield  {journal}
  {\bibinfo  {journal} {Phys. Rev. Phys. Educ. Res.}\ }\textbf {\bibinfo
  {volume} {20}},\ \bibinfo {pages} {020108} (\bibinfo {year}
  {2024})}\BibitemShut {NoStop}%
\bibitem [{\citenamefont {Borish}\ and\ \citenamefont
  {Lewandowski}(2023{\natexlab{a}})}]{borish2023seeing}%
  \BibitemOpen
  \bibfield  {author} {\bibinfo {author} {\bibfnamefont {V.}~\bibnamefont
  {Borish}}\ and\ \bibinfo {author} {\bibfnamefont {H.~J.}\ \bibnamefont
  {Lewandowski}},\ }\bibfield  {title} {\bibinfo {title} {Seeing quantum
  effects in experiments},\ }\href
  {https://doi.org/10.1103/PhysRevPhysEducRes.19.020144} {\bibfield  {journal}
  {\bibinfo  {journal} {Phys. Rev. Phys. Educ. Res.}\ }\textbf {\bibinfo
  {volume} {19}},\ \bibinfo {pages} {020144} (\bibinfo {year}
  {2023}{\natexlab{a}})}\BibitemShut {NoStop}%
\bibitem [{\citenamefont {Lukishova}(2022)}]{lukishova2022fifteen}%
  \BibitemOpen
  \bibfield  {author} {\bibinfo {author} {\bibfnamefont {S.~G.}\ \bibnamefont
  {Lukishova}},\ }\bibfield  {title} {\bibinfo {title} {Fifteen years of
  quantum optics, quantum information, and nano-optics educational facility at
  the {I}nstitute of {O}ptics, {U}niversity of {R}ochester},\ }\href
  {https://doi.org/10.1117/1.OE.61.8.081811} {\bibfield  {journal} {\bibinfo
  {journal} {Opt. Eng.}\ }\textbf {\bibinfo {volume} {61}},\ \bibinfo {pages}
  {081811} (\bibinfo {year} {2022})}\BibitemShut {NoStop}%
\bibitem [{\citenamefont {Borish}\ \emph {et~al.}(2022)\citenamefont {Borish},
  \citenamefont {Werth},\ and\ \citenamefont {Lewandowski}}]{borish2022seeing}%
  \BibitemOpen
  \bibfield  {author} {\bibinfo {author} {\bibfnamefont {V.}~\bibnamefont
  {Borish}}, \bibinfo {author} {\bibfnamefont {A.}~\bibnamefont {Werth}},\ and\
  \bibinfo {author} {\bibfnamefont {H.~J.}\ \bibnamefont {Lewandowski}},\
  }\bibfield  {title} {\bibinfo {title} {Seeing quantum mechanics: {T}he role
  of quantum experiments},\ }in\ \href
  {https://doi.org/10.1119/perc.2022.pr.Borish} {\emph {\bibinfo {booktitle}
  {Proceedings of the 2022 Physics Education Research Conference}}}\ (\bibinfo
  {year} {2022})\BibitemShut {NoStop}%
\bibitem [{\citenamefont {Lukishova}(2024)}]{lukishova2024teach}%
  \BibitemOpen
  \bibfield  {author} {\bibinfo {author} {\bibfnamefont {S.~G.}\ \bibnamefont
  {Lukishova}},\ }\bibfield  {title} {\bibinfo {title} {How to teach quantum in
  the age of the second quantum revolution: overview of the current state of
  the art},\ }\href {https://doi.org/10.1117/12.3027899} {\bibfield  {journal}
  {\bibinfo  {journal} {Optics Education and Outreach VIII}\ }\textbf {\bibinfo
  {volume} {13128}},\ \bibinfo {pages} {37} (\bibinfo {year}
  {2024})}\BibitemShut {NoStop}%
\bibitem [{\citenamefont {Piña}\ \emph {et~al.}(2024)\citenamefont {Piña},
  \citenamefont {Verostek}, \citenamefont {Boyle}, \citenamefont {Watts},
  \citenamefont {Lawler}, \citenamefont {Cacheiro}, \citenamefont {Pradeep},
  \citenamefont {West}, \citenamefont {El-Adawy}, \citenamefont {Lewandowski},\
  and\ \citenamefont {Zwickl}}]{pina2024united}%
  \BibitemOpen
  \bibfield  {author} {\bibinfo {author} {\bibfnamefont {A.}~\bibnamefont
  {Piña}}, \bibinfo {author} {\bibfnamefont {M.}~\bibnamefont {Verostek}},
  \bibinfo {author} {\bibfnamefont {B.}~\bibnamefont {Boyle}}, \bibinfo
  {author} {\bibfnamefont {E.}~\bibnamefont {Watts}}, \bibinfo {author}
  {\bibfnamefont {M.}~\bibnamefont {Lawler}}, \bibinfo {author} {\bibfnamefont
  {M.}~\bibnamefont {Cacheiro}}, \bibinfo {author} {\bibfnamefont
  {N.}~\bibnamefont {Pradeep}}, \bibinfo {author} {\bibfnamefont
  {C.}~\bibnamefont {West}}, \bibinfo {author} {\bibfnamefont {S.}~\bibnamefont
  {El-Adawy}}, \bibinfo {author} {\bibfnamefont {H.}~\bibnamefont
  {Lewandowski}},\ and\ \bibinfo {author} {\bibfnamefont {B.}~\bibnamefont
  {Zwickl}},\ }\href@noop {} {\bibinfo {title} {United {S}tates quantum
  education landscape dataset and visualizations}},\ \bibinfo {howpublished}
  {\url{https://quantumlandscape.streamlit.app/ }} (\bibinfo {year}
  {2024})\BibitemShut {NoStop}%
\bibitem [{\citenamefont {Galvez}\ \emph {et~al.}(2005)\citenamefont {Galvez},
  \citenamefont {Holbrow}, \citenamefont {Pysher}, \citenamefont {Martin},
  \citenamefont {Courtemanche}, \citenamefont {Heilig},\ and\ \citenamefont
  {Spencer}}]{galvez2005interference}%
  \BibitemOpen
  \bibfield  {author} {\bibinfo {author} {\bibfnamefont {E.~J.}\ \bibnamefont
  {Galvez}}, \bibinfo {author} {\bibfnamefont {C.~H.}\ \bibnamefont {Holbrow}},
  \bibinfo {author} {\bibfnamefont {M.}~\bibnamefont {Pysher}}, \bibinfo
  {author} {\bibfnamefont {J.}~\bibnamefont {Martin}}, \bibinfo {author}
  {\bibfnamefont {N.}~\bibnamefont {Courtemanche}}, \bibinfo {author}
  {\bibfnamefont {L.}~\bibnamefont {Heilig}},\ and\ \bibinfo {author}
  {\bibfnamefont {J.}~\bibnamefont {Spencer}},\ }\bibfield  {title} {\bibinfo
  {title} {Interference with correlated photons: {F}ive quantum mechanics
  experiments for undergraduates},\ }\href {https://doi.org/10.1119/1.1796811}
  {\bibfield  {journal} {\bibinfo  {journal} {Am. J. Phys.}\ }\textbf {\bibinfo
  {volume} {73}},\ \bibinfo {pages} {127} (\bibinfo {year} {2005})}\BibitemShut
  {NoStop}%
\bibitem [{\citenamefont {Beck}(2012)}]{beck2012quantum}%
  \BibitemOpen
  \bibfield  {author} {\bibinfo {author} {\bibfnamefont {M.}~\bibnamefont
  {Beck}},\ }\href@noop {} {\emph {\bibinfo {title} {Quantum mechanics:
  {T}heory and experiment}}}\ (\bibinfo  {publisher} {Oxford University
  Press},\ \bibinfo {year} {2012})\BibitemShut {NoStop}%
\bibitem [{\citenamefont {Borish}\ and\ \citenamefont
  {Lewandowski}(2023{\natexlab{b}})}]{borish2023implementation}%
  \BibitemOpen
  \bibfield  {author} {\bibinfo {author} {\bibfnamefont {V.}~\bibnamefont
  {Borish}}\ and\ \bibinfo {author} {\bibfnamefont {H.~J.}\ \bibnamefont
  {Lewandowski}},\ }\bibfield  {title} {\bibinfo {title} {Implementation and
  goals of quantum optics experiments in undergraduate instructional labs},\
  }\href {https://doi.org/10.1103/PhysRevPhysEducRes.19.010117} {\bibfield
  {journal} {\bibinfo  {journal} {Phys. Rev. Phys. Educ. Res.}\ }\textbf
  {\bibinfo {volume} {19}},\ \bibinfo {pages} {010117} (\bibinfo {year}
  {2023}{\natexlab{b}})}\BibitemShut {NoStop}%
\bibitem [{\citenamefont {Oliver}\ \emph {et~al.}(2024)\citenamefont {Oliver},
  \citenamefont {Borish}, \citenamefont {Wilcox},\ and\ \citenamefont
  {Lewandowski}}]{oliver2024education}%
  \BibitemOpen
  \bibfield  {author} {\bibinfo {author} {\bibfnamefont {K.~A.}\ \bibnamefont
  {Oliver}}, \bibinfo {author} {\bibfnamefont {V.}~\bibnamefont {Borish}},
  \bibinfo {author} {\bibfnamefont {B.~R.}\ \bibnamefont {Wilcox}},\ and\
  \bibinfo {author} {\bibfnamefont {H.}~\bibnamefont {Lewandowski}},\
  }\bibfield  {title} {\bibinfo {title} {Education for expanding the quantum
  workforce: Student perceptions of the quantum industry in an upper-division
  physics capstone course},\ }\href {https://arxiv.org/abs/2407.07902}
  {\bibfield  {journal} {\bibinfo  {journal} {arXiv:2407.07902}\ } (\bibinfo
  {year} {2024})}\BibitemShut {NoStop}%
\bibitem [{\citenamefont {Hunter}\ \emph {et~al.}(2007)\citenamefont {Hunter},
  \citenamefont {Laursen},\ and\ \citenamefont {Seymour}}]{hunter2007becoming}%
  \BibitemOpen
  \bibfield  {author} {\bibinfo {author} {\bibfnamefont {A.-B.}\ \bibnamefont
  {Hunter}}, \bibinfo {author} {\bibfnamefont {S.~L.}\ \bibnamefont
  {Laursen}},\ and\ \bibinfo {author} {\bibfnamefont {E.}~\bibnamefont
  {Seymour}},\ }\bibfield  {title} {\bibinfo {title} {Becoming a scientist:
  {T}he role of undergraduate research in students' cognitive, personal, and
  professional development},\ }\href {https://doi.org/10.1002/sce.20173}
  {\bibfield  {journal} {\bibinfo  {journal} {Science Education}\ }\textbf
  {\bibinfo {volume} {91}},\ \bibinfo {pages} {36} (\bibinfo {year}
  {2007})}\BibitemShut {NoStop}%
\bibitem [{\citenamefont {McKagan}\ \emph {et~al.}(2008)\citenamefont
  {McKagan}, \citenamefont {Perkins}, \citenamefont {Dubson}, \citenamefont
  {Malley}, \citenamefont {Reid}, \citenamefont {LeMaster},\ and\ \citenamefont
  {Wieman}}]{mckagan2008developing}%
  \BibitemOpen
  \bibfield  {author} {\bibinfo {author} {\bibfnamefont {S.}~\bibnamefont
  {McKagan}}, \bibinfo {author} {\bibfnamefont {K.~K.}\ \bibnamefont
  {Perkins}}, \bibinfo {author} {\bibfnamefont {M.}~\bibnamefont {Dubson}},
  \bibinfo {author} {\bibfnamefont {C.}~\bibnamefont {Malley}}, \bibinfo
  {author} {\bibfnamefont {S.}~\bibnamefont {Reid}}, \bibinfo {author}
  {\bibfnamefont {R.}~\bibnamefont {LeMaster}},\ and\ \bibinfo {author}
  {\bibfnamefont {C.}~\bibnamefont {Wieman}},\ }\bibfield  {title} {\bibinfo
  {title} {Developing and researching {PhET} simulations for teaching quantum
  mechanics},\ }\href {https://doi.org/10.1119/1.2885199} {\bibfield  {journal}
  {\bibinfo  {journal} {Am. J. Phys.}\ }\textbf {\bibinfo {volume} {76}},\
  \bibinfo {pages} {406} (\bibinfo {year} {2008})}\BibitemShut {NoStop}%
\bibitem [{\citenamefont {Ahmed}\ \emph {et~al.}(2022)\citenamefont {Ahmed},
  \citenamefont {Weidner}, \citenamefont {Jensen}, \citenamefont {Sherson},\
  and\ \citenamefont {Lewandowski}}]{ahmed2022student}%
  \BibitemOpen
  \bibfield  {author} {\bibinfo {author} {\bibfnamefont {S.~Z.}\ \bibnamefont
  {Ahmed}}, \bibinfo {author} {\bibfnamefont {C.~A.}\ \bibnamefont {Weidner}},
  \bibinfo {author} {\bibfnamefont {J.~H.}\ \bibnamefont {Jensen}}, \bibinfo
  {author} {\bibfnamefont {J.~F.}\ \bibnamefont {Sherson}},\ and\ \bibinfo
  {author} {\bibfnamefont {H.}~\bibnamefont {Lewandowski}},\ }\bibfield
  {title} {\bibinfo {title} {Student use of a quantum simulation and
  visualization tool},\ }\href {https://doi.org/10.1088/1361-6404/ac93c7}
  {\bibfield  {journal} {\bibinfo  {journal} {Eur. J. Phys.}\ }\textbf
  {\bibinfo {volume} {43}},\ \bibinfo {pages} {065703} (\bibinfo {year}
  {2022})}\BibitemShut {NoStop}%
\bibitem [{\citenamefont {Kohnle}\ \emph {et~al.}(2015)\citenamefont {Kohnle},
  \citenamefont {Baily}, \citenamefont {Campbell}, \citenamefont {Korolkova},\
  and\ \citenamefont {Paetkau}}]{kohnle2015enhancing}%
  \BibitemOpen
  \bibfield  {author} {\bibinfo {author} {\bibfnamefont {A.}~\bibnamefont
  {Kohnle}}, \bibinfo {author} {\bibfnamefont {C.}~\bibnamefont {Baily}},
  \bibinfo {author} {\bibfnamefont {A.}~\bibnamefont {Campbell}}, \bibinfo
  {author} {\bibfnamefont {N.}~\bibnamefont {Korolkova}},\ and\ \bibinfo
  {author} {\bibfnamefont {M.~J.}\ \bibnamefont {Paetkau}},\ }\bibfield
  {title} {\bibinfo {title} {Enhancing student learning of two-level quantum
  systems with interactive simulations},\ }\href
  {https://doi.org/10.1119/1.4913786} {\bibfield  {journal} {\bibinfo
  {journal} {Am. J. Phys.}\ }\textbf {\bibinfo {volume} {83}},\ \bibinfo
  {pages} {560} (\bibinfo {year} {2015})}\BibitemShut {NoStop}%
\bibitem [{\citenamefont {Malgieri}\ \emph {et~al.}(2014)\citenamefont
  {Malgieri}, \citenamefont {Onorato},\ and\ \citenamefont
  {De~Ambrosis}}]{malgieri2014teaching}%
  \BibitemOpen
  \bibfield  {author} {\bibinfo {author} {\bibfnamefont {M.}~\bibnamefont
  {Malgieri}}, \bibinfo {author} {\bibfnamefont {P.}~\bibnamefont {Onorato}},\
  and\ \bibinfo {author} {\bibfnamefont {A.}~\bibnamefont {De~Ambrosis}},\
  }\bibfield  {title} {\bibinfo {title} {Teaching quantum physics by the sum
  over paths approach and {G}eo{G}ebra simulations},\ }\href
  {https://doi.org/10.1088/0143-0807/35/5/055024} {\bibfield  {journal}
  {\bibinfo  {journal} {Eur. J. Phys.}\ }\textbf {\bibinfo {volume} {35}},\
  \bibinfo {pages} {055024} (\bibinfo {year} {2014})}\BibitemShut {NoStop}%
\bibitem [{\citenamefont {Marshman}\ and\ \citenamefont
  {Singh}(2022)}]{marshman2022quilts}%
  \BibitemOpen
  \bibfield  {author} {\bibinfo {author} {\bibfnamefont {E.}~\bibnamefont
  {Marshman}}\ and\ \bibinfo {author} {\bibfnamefont {C.}~\bibnamefont
  {Singh}},\ }\bibfield  {title} {\bibinfo {title} {{QuILTs}: Validated
  teaching--learning sequences for helping students learn quantum mechanics},\
  }in\ \href {https://doi.org/10.1007/978-3-031-06193-6_2} {\emph {\bibinfo
  {booktitle} {Physics Teacher Education: What Matters?}}}\ (\bibinfo
  {publisher} {Springer},\ \bibinfo {year} {2022})\ pp.\ \bibinfo {pages}
  {15--35}\BibitemShut {NoStop}%
\bibitem [{\citenamefont {Marshman}\ and\ \citenamefont
  {Singh}(2016)}]{marshman2016interactive}%
  \BibitemOpen
  \bibfield  {author} {\bibinfo {author} {\bibfnamefont {E.}~\bibnamefont
  {Marshman}}\ and\ \bibinfo {author} {\bibfnamefont {C.}~\bibnamefont
  {Singh}},\ }\bibfield  {title} {\bibinfo {title} {Interactive tutorial to
  improve student understanding of single photon experiments involving a
  {M}ach--{Z}ehnder interferometer},\ }\href
  {https://doi.org/10.1088/0143-0807/37/2/024001} {\bibfield  {journal}
  {\bibinfo  {journal} {Eur. J. Phys.}\ }\textbf {\bibinfo {volume} {37}},\
  \bibinfo {pages} {024001} (\bibinfo {year} {2016})}\BibitemShut {NoStop}%
\bibitem [{\citenamefont {La~Cour}\ \emph {et~al.}(2021)\citenamefont
  {La~Cour}, \citenamefont {Maynard}, \citenamefont {Shroff}, \citenamefont
  {Ko},\ and\ \citenamefont {Ellis}}]{la2021virtual}%
  \BibitemOpen
  \bibfield  {author} {\bibinfo {author} {\bibfnamefont {B.~R.}\ \bibnamefont
  {La~Cour}}, \bibinfo {author} {\bibfnamefont {M.}~\bibnamefont {Maynard}},
  \bibinfo {author} {\bibfnamefont {P.}~\bibnamefont {Shroff}}, \bibinfo
  {author} {\bibfnamefont {G.}~\bibnamefont {Ko}},\ and\ \bibinfo {author}
  {\bibfnamefont {E.}~\bibnamefont {Ellis}},\ }\bibfield  {title} {\bibinfo
  {title} {The virtual quantum optics laboratory},\ }\href
  {https://doi.org/10.48550/arXiv.2105.07300} {\bibfield  {journal} {\bibinfo
  {journal} {arXiv:2105.07300}\ } (\bibinfo {year} {2021})}\BibitemShut
  {NoStop}%
\bibitem [{\citenamefont {Migda{\l}}\ \emph {et~al.}(2022)\citenamefont
  {Migda{\l}}, \citenamefont {Jankiewicz}, \citenamefont {Grabarz},
  \citenamefont {Decaroli},\ and\ \citenamefont
  {Cochin}}]{migdal2022visualizing}%
  \BibitemOpen
  \bibfield  {author} {\bibinfo {author} {\bibfnamefont {P.}~\bibnamefont
  {Migda{\l}}}, \bibinfo {author} {\bibfnamefont {K.}~\bibnamefont
  {Jankiewicz}}, \bibinfo {author} {\bibfnamefont {P.}~\bibnamefont {Grabarz}},
  \bibinfo {author} {\bibfnamefont {C.}~\bibnamefont {Decaroli}},\ and\
  \bibinfo {author} {\bibfnamefont {P.}~\bibnamefont {Cochin}},\ }\bibfield
  {title} {\bibinfo {title} {Visualizing quantum mechanics in an interactive
  simulation--{V}irtual {L}ab by {Q}uantum {F}lytrap},\ }\href
  {https://doi.org/10.1117/1.OE.61.8.081808} {\bibfield  {journal} {\bibinfo
  {journal} {Opt. Eng.}\ }\textbf {\bibinfo {volume} {61}} (\bibinfo {year}
  {2022})}\BibitemShut {NoStop}%
\bibitem [{\citenamefont {Waitzmann}\ \emph {et~al.}(2024)\citenamefont
  {Waitzmann}, \citenamefont {Scholz},\ and\ \citenamefont
  {Wessnigk}}]{waitzmann2024testing}%
  \BibitemOpen
  \bibfield  {author} {\bibinfo {author} {\bibfnamefont {M.}~\bibnamefont
  {Waitzmann}}, \bibinfo {author} {\bibfnamefont {R.}~\bibnamefont {Scholz}},\
  and\ \bibinfo {author} {\bibfnamefont {S.}~\bibnamefont {Wessnigk}},\
  }\bibfield  {title} {\bibinfo {title} {Testing quantum reasoning: Developing,
  validating, and application of a questionnaire},\ }\href
  {https://doi.org/10.1103/PhysRevPhysEducRes.20.010122} {\bibfield  {journal}
  {\bibinfo  {journal} {Phys. Rev. Phys. Educ. Res.}\ }\textbf {\bibinfo
  {volume} {20}},\ \bibinfo {pages} {010122} (\bibinfo {year}
  {2024})}\BibitemShut {NoStop}%
\bibitem [{\citenamefont {Bronner}\ \emph {et~al.}(2009)\citenamefont
  {Bronner}, \citenamefont {Strunz}, \citenamefont {Silberhorn},\ and\
  \citenamefont {Meyn}}]{bronner2009interactive}%
  \BibitemOpen
  \bibfield  {author} {\bibinfo {author} {\bibfnamefont {P.}~\bibnamefont
  {Bronner}}, \bibinfo {author} {\bibfnamefont {A.}~\bibnamefont {Strunz}},
  \bibinfo {author} {\bibfnamefont {C.}~\bibnamefont {Silberhorn}},\ and\
  \bibinfo {author} {\bibfnamefont {J.-P.}\ \bibnamefont {Meyn}},\ }\bibfield
  {title} {\bibinfo {title} {Interactive screen experiments with single
  photons},\ }\href {http://doi.org/10.1088/0143-0807/30/2/014} {\bibfield
  {journal} {\bibinfo  {journal} {Eur. J. Phys.}\ }\textbf {\bibinfo {volume}
  {30}},\ \bibinfo {pages} {345} (\bibinfo {year} {2009})}\BibitemShut
  {NoStop}%
\bibitem [{\citenamefont {Bitzenbauer}(2021)}]{bitzenbauer2021effect}%
  \BibitemOpen
  \bibfield  {author} {\bibinfo {author} {\bibfnamefont {P.}~\bibnamefont
  {Bitzenbauer}},\ }\bibfield  {title} {\bibinfo {title} {Effect of an
  introductory quantum physics course using experiments with heralded photons
  on preuniversity students' conceptions about quantum physics},\ }\href
  {https://doi.org/10.1103/PhysRevPhysEducRes.17.020103} {\bibfield  {journal}
  {\bibinfo  {journal} {Phys. Rev. Phys. Educ. Res.}\ }\textbf {\bibinfo
  {volume} {17}},\ \bibinfo {pages} {020103} (\bibinfo {year}
  {2021})}\BibitemShut {NoStop}%
\bibitem [{\citenamefont {Galvez}(2021)}]{galvez2021remote}%
  \BibitemOpen
  \bibfield  {author} {\bibinfo {author} {\bibfnamefont {E.~J.}\ \bibnamefont
  {Galvez}},\ }\bibfield  {title} {\bibinfo {title} {Remote quantum optics
  labs},\ }in\ \href {https://doi.org/10.1117/12.2584597} {\emph {\bibinfo
  {booktitle} {Complex Light and Optical Forces XV}}},\ Vol.\ \bibinfo {volume}
  {11701}\ (\bibinfo {organization} {SPIE},\ \bibinfo {year}
  {2021})\BibitemShut {NoStop}%
\bibitem [{\citenamefont {He}(2021)}]{UCSBremoteLabs}%
  \BibitemOpen
  \bibfield  {author} {\bibinfo {author} {\bibfnamefont {C.}~\bibnamefont
  {He}},\ }\bibfield  {title} {\bibinfo {title} {\$10,000 and six months later:
  How the physics department automated lab courses for remote learning},\
  }\href@noop {} {\bibfield  {journal} {\bibinfo  {journal} {UC Santa Barbara
  Daily Nexus}\ } (\bibinfo {year} {2021})},\ \bibinfo {note}
  {\url{https://dailynexus.com/2021-01-03/10000-and-six-months-later-how-the-physics-department-automated-lab-courses-for-remote-learning/}}\BibitemShut
  {NoStop}%
\bibitem [{Ion()}]{IonQquantumCloud}%
  \BibitemOpen
  \href@noop {} {}\bibinfo {note} {IonQ Quantum Cloud,
  \url{https://ionq.com/quantum-cloud}}\BibitemShut {NoStop}%
\bibitem [{QuE()}]{QuEra}%
  \BibitemOpen
  \href@noop {} {}\bibinfo {note} {QuEra Aquila,
  \url{https://www.quera.com/aquila}}\BibitemShut {NoStop}%
\bibitem [{qub()}]{qubitByQubit}%
  \BibitemOpen
  \href@noop {} {}\bibinfo {note} {Qubit by Qubit,
  \url{https://www.qubitbyqubit.org/programs}}\BibitemShut {NoStop}%
\bibitem [{\citenamefont {Tappert}\ \emph {et~al.}(2019)\citenamefont
  {Tappert}, \citenamefont {Frank}, \citenamefont {Barabasi}, \citenamefont
  {Leider}, \citenamefont {Evans},\ and\ \citenamefont
  {Westfall}}]{tappert2019experience}%
  \BibitemOpen
  \bibfield  {author} {\bibinfo {author} {\bibfnamefont {C.~C.}\ \bibnamefont
  {Tappert}}, \bibinfo {author} {\bibfnamefont {R.~I.}\ \bibnamefont {Frank}},
  \bibinfo {author} {\bibfnamefont {I.}~\bibnamefont {Barabasi}}, \bibinfo
  {author} {\bibfnamefont {A.~M.}\ \bibnamefont {Leider}}, \bibinfo {author}
  {\bibfnamefont {D.}~\bibnamefont {Evans}},\ and\ \bibinfo {author}
  {\bibfnamefont {L.}~\bibnamefont {Westfall}},\ }\bibfield  {title} {\bibinfo
  {title} {Experience teaching quantum computing.},\ }\href
  {https://eric.ed.gov/?id=ED597112} {\bibfield  {journal} {\bibinfo  {journal}
  {Association Supporting Computer Users in Education}\ } (\bibinfo {year}
  {2019})}\BibitemShut {NoStop}%
\bibitem [{Qis()}]{QiskitSummerSchool}%
  \BibitemOpen
  \href@noop {} {}\bibinfo {note} {Qiskit Global Summer School 2023,
  \url{https://github.com/qiskit-community/qgss-2023/tree/main}}\BibitemShut
  {NoStop}%
\bibitem [{\citenamefont {Singh}\ \emph {et~al.}(2021)\citenamefont {Singh},
  \citenamefont {Asfaw},\ and\ \citenamefont {Levy}}]{singh2021preparing}%
  \BibitemOpen
  \bibfield  {author} {\bibinfo {author} {\bibfnamefont {C.}~\bibnamefont
  {Singh}}, \bibinfo {author} {\bibfnamefont {A.}~\bibnamefont {Asfaw}},\ and\
  \bibinfo {author} {\bibfnamefont {J.}~\bibnamefont {Levy}},\ }\bibfield
  {title} {\bibinfo {title} {Preparing students to be leaders of the quantum
  information revolution},\ }\bibfield  {journal} {\bibinfo  {journal} {Physics
  Today}\ }\href {https://doi.org/10.1063/PT.6.5.20210927a}
  {10.1063/PT.6.5.20210927a} (\bibinfo {year} {2021})\BibitemShut {NoStop}%
\bibitem [{\citenamefont {Brody}\ and\ \citenamefont
  {Avram}(2023)}]{brody2023testing}%
  \BibitemOpen
  \bibfield  {author} {\bibinfo {author} {\bibfnamefont {J.}~\bibnamefont
  {Brody}}\ and\ \bibinfo {author} {\bibfnamefont {R.}~\bibnamefont {Avram}},\
  }\bibfield  {title} {\bibinfo {title} {Testing a {B}ell inequality with a
  remote quantum processor},\ }\href {https://doi.org/10.1119/5.0069073}
  {\bibfield  {journal} {\bibinfo  {journal} {The Physics Teacher}\ }\textbf
  {\bibinfo {volume} {61}},\ \bibinfo {pages} {218} (\bibinfo {year}
  {2023})}\BibitemShut {NoStop}%
\bibitem [{\citenamefont {Tran}\ \emph {et~al.}(2022)\citenamefont {Tran},
  \citenamefont {Narong},\ and\ \citenamefont {Cooper}}]{tran2022modeling}%
  \BibitemOpen
  \bibfield  {author} {\bibinfo {author} {\bibfnamefont {C.}~\bibnamefont
  {Tran}}, \bibinfo {author} {\bibfnamefont {T.~N.}\ \bibnamefont {Narong}},\
  and\ \bibinfo {author} {\bibfnamefont {E.~S.}\ \bibnamefont {Cooper}},\
  }\bibfield  {title} {\bibinfo {title} {Modeling quantum enhanced sensing on a
  quantum computer},\ }\href {https://doi.org/10.48550/arXiv.2209.08187}
  {\bibfield  {journal} {\bibinfo  {journal} {arXiv:2209.08187}\ } (\bibinfo
  {year} {2022})}\BibitemShut {NoStop}%
\bibitem [{\citenamefont {Pitzevskii}\ and\ \citenamefont
  {Stringari}(2003)}]{pitaevskii2003bose}%
  \BibitemOpen
  \bibfield  {author} {\bibinfo {author} {\bibfnamefont {L.}~\bibnamefont
  {Pitzevskii}}\ and\ \bibinfo {author} {\bibfnamefont {S.}~\bibnamefont
  {Stringari}},\ }\href@noop {} {\emph {\bibinfo {title} {Bose-{E}instein
  Condensation}}}\ (\bibinfo  {publisher} {Oxford University Press},\ \bibinfo
  {year} {2003})\BibitemShut {NoStop}%
\bibitem [{\citenamefont {Salim}\ \emph {et~al.}(2013)\citenamefont {Salim},
  \citenamefont {Caliga}, \citenamefont {Pfeiffer},\ and\ \citenamefont
  {Anderson}}]{salim2013high}%
  \BibitemOpen
  \bibfield  {author} {\bibinfo {author} {\bibfnamefont {E.~A.}\ \bibnamefont
  {Salim}}, \bibinfo {author} {\bibfnamefont {S.~C.}\ \bibnamefont {Caliga}},
  \bibinfo {author} {\bibfnamefont {J.~B.}\ \bibnamefont {Pfeiffer}},\ and\
  \bibinfo {author} {\bibfnamefont {D.~Z.}\ \bibnamefont {Anderson}},\
  }\bibfield  {title} {\bibinfo {title} {High resolution imaging and optical
  control of {B}ose-{E}instein condensates in an atom chip magnetic trap},\
  }\bibfield  {journal} {\bibinfo  {journal} {Applied Physics Letters}\
  }\textbf {\bibinfo {volume} {102}},\ \href
  {https://doi.org/10.1063/1.4793522} {10.1063/1.4793522} (\bibinfo {year}
  {2013})\BibitemShut {NoStop}%
\bibitem [{\citenamefont {Lewandowski}\ \emph {et~al.}(2003)\citenamefont
  {Lewandowski}, \citenamefont {Harber}, \citenamefont {Whitaker},\ and\
  \citenamefont {Cornell}}]{lewandowski2003simplified}%
  \BibitemOpen
  \bibfield  {author} {\bibinfo {author} {\bibfnamefont {H.~J.}\ \bibnamefont
  {Lewandowski}}, \bibinfo {author} {\bibfnamefont {D.}~\bibnamefont {Harber}},
  \bibinfo {author} {\bibfnamefont {D.~L.}\ \bibnamefont {Whitaker}},\ and\
  \bibinfo {author} {\bibfnamefont {E.~A.}\ \bibnamefont {Cornell}},\
  }\bibfield  {title} {\bibinfo {title} {Simplified system for creating a
  {B}ose--{E}instein condensate},\ }\href
  {https://doi.org/10.1023/A:1024800600621} {\bibfield  {journal} {\bibinfo
  {journal} {Journal of low temperature physics}\ }\textbf {\bibinfo {volume}
  {132}},\ \bibinfo {pages} {309} (\bibinfo {year} {2003})}\BibitemShut
  {NoStop}%
\bibitem [{\citenamefont {Inguscio}\ \emph {et~al.}(1999)\citenamefont
  {Inguscio}, \citenamefont {Stringari},\ and\ \citenamefont
  {Wieman}}]{inguscio1999bose}%
  \BibitemOpen
  \bibfield  {author} {\bibinfo {author} {\bibfnamefont {M.}~\bibnamefont
  {Inguscio}}, \bibinfo {author} {\bibfnamefont {S.}~\bibnamefont
  {Stringari}},\ and\ \bibinfo {author} {\bibfnamefont {C.}~\bibnamefont
  {Wieman}},\ }\href@noop {} {\emph {\bibinfo {title} {Bose-{E}instein
  condensation in atomic gases}}},\ Vol.\ \bibinfo {volume} {140}\ (\bibinfo
  {publisher} {IOS Press},\ \bibinfo {year} {1999})\BibitemShut {NoStop}%
\bibitem [{alp()}]{alphaWebsite}%
  \BibitemOpen
  \href@noop {} {}\bibinfo {note} {Advanced Laboratory Physics Association,
  \url{https://advlab.org/}}\BibitemShut {NoStop}%
\bibitem [{SM()}]{SM}%
  \BibitemOpen
  \href@noop {} {}\bibinfo {note} {See Supplemental Material at [URL will be
  inserted by publisher] for details about the data sources and data
  analysis.}\BibitemShut {Stop}%
\bibitem [{\citenamefont {Holmes}\ and\ \citenamefont
  {Smith}(2023)}]{holmes2023strategies}%
  \BibitemOpen
  \bibfield  {author} {\bibinfo {author} {\bibfnamefont {N.~G.}\ \bibnamefont
  {Holmes}}\ and\ \bibinfo {author} {\bibfnamefont {E.~M.}\ \bibnamefont
  {Smith}},\ }\bibfield  {title} {\bibinfo {title} {Instructional strategies
  that foster experimental physics skills},\ }in\ \href
  {https://doi.org/10.1063/9780735425477_018} {\emph {\bibinfo {booktitle} {The
  International Handbook of Physics Education Research: Learning Physics}}}\
  (\bibinfo  {publisher} {AIP Publishing},\ \bibinfo {year} {2023})\ pp.\
  \bibinfo {pages} {18--1 -- 18--20}\BibitemShut {NoStop}%
\bibitem [{\citenamefont {Simon}\ and\ \citenamefont
  {Taylor}(2009)}]{simon2009value}%
  \BibitemOpen
  \bibfield  {author} {\bibinfo {author} {\bibfnamefont {B.}~\bibnamefont
  {Simon}}\ and\ \bibinfo {author} {\bibfnamefont {J.}~\bibnamefont {Taylor}},\
  }\bibfield  {title} {\bibinfo {title} {What is the value of course-specific
  learning goals?},\ }\href
  {https://sei.ubc.ca/bitstream/seima/2106/1/Simon_Taylor_ValueOfCourseSpecificLG.pdf}
  {\bibfield  {journal} {\bibinfo  {journal} {Journal of College Science
  Teaching}\ } (\bibinfo {year} {2009})}\BibitemShut {NoStop}%
\bibitem [{Lew()}]{LewGroupWebsite}%
  \BibitemOpen
  \href@noop {} {}\bibinfo {note} {The Lewandowski Group: Quantum Education,
  \url{https://jila.colorado.edu/lewandowski/research/quantum-education-1}}\BibitemShut
  {NoStop}%
\bibitem [{\citenamefont {Charters}(2003)}]{charters2003use}%
  \BibitemOpen
  \bibfield  {author} {\bibinfo {author} {\bibfnamefont {E.}~\bibnamefont
  {Charters}},\ }\bibfield  {title} {\bibinfo {title} {The use of think-aloud
  methods in qualitative research: {A}n introduction to think-aloud methods},\
  }\bibfield  {journal} {\bibinfo  {journal} {Brock Education Journal}\
  }\textbf {\bibinfo {volume} {12}},\ \href
  {https://doi.org/10.26522/brocked.v12i2.38} {10.26522/brocked.v12i2.38}
  (\bibinfo {year} {2003})\BibitemShut {NoStop}%
\bibitem [{\citenamefont {Braun}\ and\ \citenamefont
  {Clarke}(2006)}]{braun2006using}%
  \BibitemOpen
  \bibfield  {author} {\bibinfo {author} {\bibfnamefont {V.}~\bibnamefont
  {Braun}}\ and\ \bibinfo {author} {\bibfnamefont {V.}~\bibnamefont {Clarke}},\
  }\bibfield  {title} {\bibinfo {title} {Using thematic analysis in
  psychology},\ }\href {https://doi.org/10.1191/1478088706qp063oa} {\bibfield
  {journal} {\bibinfo  {journal} {Qualitative Research in Psychology}\ }\textbf
  {\bibinfo {volume} {3}},\ \bibinfo {pages} {77} (\bibinfo {year}
  {2006})}\BibitemShut {NoStop}%
\bibitem [{\citenamefont {Merriam}\ and\ \citenamefont
  {Tisdell}(2015)}]{merriam2015qualitative}%
  \BibitemOpen
  \bibfield  {author} {\bibinfo {author} {\bibfnamefont {S.~B.}\ \bibnamefont
  {Merriam}}\ and\ \bibinfo {author} {\bibfnamefont {E.~J.}\ \bibnamefont
  {Tisdell}},\ }\href@noop {} {\emph {\bibinfo {title} {Qualitative research: A
  guide to design and implementation}}}\ (\bibinfo  {publisher} {John Wiley \&
  Sons},\ \bibinfo {year} {2015})\BibitemShut {NoStop}%
\bibitem [{\citenamefont {Eisenhart}(2009)}]{eisenhart2009generalization}%
  \BibitemOpen
  \bibfield  {author} {\bibinfo {author} {\bibfnamefont {M.}~\bibnamefont
  {Eisenhart}},\ }\bibfield  {title} {\bibinfo {title} {Generalization from
  qualitative inquiry},\ }in\ \href@noop {} {\emph {\bibinfo {booktitle}
  {Generalizing from educational research: Beyond qualitative and quantitative
  polarization}}}\ (\bibinfo {year} {2009})\ pp.\ \bibinfo {pages}
  {51--66}\BibitemShut {NoStop}%
\bibitem [{\citenamefont {Robertson}\ \emph {et~al.}(2018)\citenamefont
  {Robertson}, \citenamefont {McKagan},\ and\ \citenamefont
  {Scherr}}]{robertson2018selection}%
  \BibitemOpen
  \bibfield  {author} {\bibinfo {author} {\bibfnamefont {A.~D.}\ \bibnamefont
  {Robertson}}, \bibinfo {author} {\bibfnamefont {S.~B.}\ \bibnamefont
  {McKagan}},\ and\ \bibinfo {author} {\bibfnamefont {R.~E.}\ \bibnamefont
  {Scherr}},\ }\bibfield  {title} {\bibinfo {title} {Selection, generalization,
  and theories of cause in physics education research: Connecting paradigms and
  practices},\ }\bibfield  {booktitle} {\emph {\bibinfo {booktitle} {Getting
  Started in PER}},\ }\href
  {http://www.per-central.org/items/detail.cfm?ID=14727} {\ \textbf {\bibinfo
  {volume} {2}} (\bibinfo {year} {2018})}\BibitemShut {NoStop}%
\bibitem [{\citenamefont {Wilcox}\ and\ \citenamefont
  {Lewandowski}(2016)}]{wilcox2016open}%
  \BibitemOpen
  \bibfield  {author} {\bibinfo {author} {\bibfnamefont {B.~R.}\ \bibnamefont
  {Wilcox}}\ and\ \bibinfo {author} {\bibfnamefont {H.~J.}\ \bibnamefont
  {Lewandowski}},\ }\bibfield  {title} {\bibinfo {title} {Open-ended versus
  guided laboratory activities: {I}mpact on students' beliefs about
  experimental physics},\ }\href
  {https://doi.org/10.1103/PhysRevPhysEducRes.12.020132} {\bibfield  {journal}
  {\bibinfo  {journal} {Phys. Rev. Phys. Educ. Res.}\ }\textbf {\bibinfo
  {volume} {12}},\ \bibinfo {pages} {020132} (\bibinfo {year}
  {2016})}\BibitemShut {NoStop}%
\bibitem [{\citenamefont {Smith}\ and\ \citenamefont
  {Holmes}(2021)}]{smith2021best}%
  \BibitemOpen
  \bibfield  {author} {\bibinfo {author} {\bibfnamefont {E.~M.}\ \bibnamefont
  {Smith}}\ and\ \bibinfo {author} {\bibfnamefont {N.~G.}\ \bibnamefont
  {Holmes}},\ }\bibfield  {title} {\bibinfo {title} {Best practice for
  instructional labs},\ }\href
  {https://www.nature.com/articles/s41567-021-01256-6} {\bibfield  {journal}
  {\bibinfo  {journal} {Nature Physics}\ }\textbf {\bibinfo {volume} {17}},\
  \bibinfo {pages} {662} (\bibinfo {year} {2021})}\BibitemShut {NoStop}%
\bibitem [{\citenamefont {Liu}\ and\ \citenamefont
  {Lewandowski}(2024)}]{liu2024correlation}%
  \BibitemOpen
  \bibfield  {author} {\bibinfo {author} {\bibfnamefont {Q.}~\bibnamefont
  {Liu}}\ and\ \bibinfo {author} {\bibfnamefont {H.~J.}\ \bibnamefont
  {Lewandowski}},\ }\bibfield  {title} {\bibinfo {title} {Correlation of
  open-ended activities in laboratory courses with students' views of
  experimental physics},\ }in\ \href {https://doi.org/10.1119/perc.2024.pr.Liu}
  {\emph {\bibinfo {booktitle} {2024 PERC Proceedings}}}\ (\bibinfo {year}
  {2024})\BibitemShut {NoStop}%
\bibitem [{\citenamefont {Barnes}\ \emph {et~al.}(2024)\citenamefont {Barnes},
  \citenamefont {Bennett}, \citenamefont {Boltasseva}, \citenamefont {Borish},
  \citenamefont {Brown}, \citenamefont {Carr}, \citenamefont {Easton},
  \citenamefont {Economou}, \citenamefont {Edwards}, \citenamefont
  {Finkelstein} \emph {et~al.}}]{barnes2024outcomes}%
  \BibitemOpen
  \bibfield  {author} {\bibinfo {author} {\bibfnamefont {E.}~\bibnamefont
  {Barnes}}, \bibinfo {author} {\bibfnamefont {M.~B.}\ \bibnamefont {Bennett}},
  \bibinfo {author} {\bibfnamefont {A.}~\bibnamefont {Boltasseva}}, \bibinfo
  {author} {\bibfnamefont {V.}~\bibnamefont {Borish}}, \bibinfo {author}
  {\bibfnamefont {B.}~\bibnamefont {Brown}}, \bibinfo {author} {\bibfnamefont
  {L.~D.}\ \bibnamefont {Carr}}, \bibinfo {author} {\bibfnamefont {E.~W.}\
  \bibnamefont {Easton}}, \bibinfo {author} {\bibfnamefont {S.~E.}\
  \bibnamefont {Economou}}, \bibinfo {author} {\bibfnamefont {E.~E.}\
  \bibnamefont {Edwards}}, \bibinfo {author} {\bibfnamefont {N.~D.}\
  \bibnamefont {Finkelstein}}, \emph {et~al.},\ }\bibfield  {title} {\bibinfo
  {title} {Outcomes from a workshop on a national center for quantum
  education},\ }\href {https://arxiv.org/abs/2410.23460} {\bibfield  {journal}
  {\bibinfo  {journal} {arXiv:2410.23460}\ } (\bibinfo {year}
  {2024})}\BibitemShut {NoStop}%
\bibitem [{oqt({\natexlab{b}})}]{oqtantDemoNotebooks}%
  \BibitemOpen
  \href@noop {} {} ({\natexlab{b}}),\ \bibinfo {note} {{O}qtant example
  notebooks,
  \url{https://oqtant-docs.infleqtion.com/examples/hello_world/}}\BibitemShut
  {NoStop}%
\bibitem [{\citenamefont {Dounas-Frazer}\ and\ \citenamefont
  {Lewandowski}(2016)}]{dounas2016nothing}%
  \BibitemOpen
  \bibfield  {author} {\bibinfo {author} {\bibfnamefont {D.}~\bibnamefont
  {Dounas-Frazer}}\ and\ \bibinfo {author} {\bibfnamefont {H.~J.}\ \bibnamefont
  {Lewandowski}},\ }\bibfield  {title} {\bibinfo {title} {Nothing works the
  first time: {A}n expert experimental physics epistemology},\ }in\ \href
  {https://doi.org/10.1119/perc.2016.pr.020} {\emph {\bibinfo {booktitle}
  {Physics Education Research Conference 2016}}},\ \bibinfo {series and number}
  {PER Conference}\ (\bibinfo {address} {Sacramento, CA},\ \bibinfo {year}
  {2016})\ pp.\ \bibinfo {pages} {100--103}\BibitemShut {NoStop}%
\bibitem [{\citenamefont {Henige}(2011)}]{henige2011undergraduate}%
  \BibitemOpen
  \bibfield  {author} {\bibinfo {author} {\bibfnamefont {K.}~\bibnamefont
  {Henige}},\ }\bibfield  {title} {\bibinfo {title} {Undergraduate student
  attitudes and perceptions toward low-and high-level inquiry exercise
  physiology teaching laboratory experiences},\ }\href
  {https://doi.org/10.1152/advan.00086.2010} {\bibfield  {journal} {\bibinfo
  {journal} {Advances in physiology education}\ }\textbf {\bibinfo {volume}
  {35}},\ \bibinfo {pages} {197} (\bibinfo {year} {2011})}\BibitemShut
  {NoStop}%
\bibitem [{\citenamefont {Owens}\ \emph {et~al.}(2020)\citenamefont {Owens},
  \citenamefont {Sadler}, \citenamefont {Barlow},\ and\ \citenamefont
  {Smith-Walters}}]{owens2020student}%
  \BibitemOpen
  \bibfield  {author} {\bibinfo {author} {\bibfnamefont {D.~C.}\ \bibnamefont
  {Owens}}, \bibinfo {author} {\bibfnamefont {T.~D.}\ \bibnamefont {Sadler}},
  \bibinfo {author} {\bibfnamefont {A.~T.}\ \bibnamefont {Barlow}},\ and\
  \bibinfo {author} {\bibfnamefont {C.}~\bibnamefont {Smith-Walters}},\
  }\bibfield  {title} {\bibinfo {title} {Student motivation from and resistance
  to active learning rooted in essential science practices},\ }\href
  {https://doi.org/10.1007/s11165-017-9688-1} {\bibfield  {journal} {\bibinfo
  {journal} {Research in Science Education}\ }\textbf {\bibinfo {volume}
  {50}},\ \bibinfo {pages} {253} (\bibinfo {year} {2020})}\BibitemShut
  {NoStop}%
\bibitem [{\citenamefont {Kalender}\ \emph {et~al.}(2021)\citenamefont
  {Kalender}, \citenamefont {Stump}, \citenamefont {Hubenig},\ and\
  \citenamefont {Holmes}}]{kalender2021restructuring}%
  \BibitemOpen
  \bibfield  {author} {\bibinfo {author} {\bibfnamefont {Z.~Y.}\ \bibnamefont
  {Kalender}}, \bibinfo {author} {\bibfnamefont {E.}~\bibnamefont {Stump}},
  \bibinfo {author} {\bibfnamefont {K.}~\bibnamefont {Hubenig}},\ and\ \bibinfo
  {author} {\bibfnamefont {N.~G.}\ \bibnamefont {Holmes}},\ }\bibfield  {title}
  {\bibinfo {title} {Restructuring physics labs to cultivate sense of student
  agency},\ }\href {https://doi.org/10.1103/PhysRevPhysEducRes.17.020128}
  {\bibfield  {journal} {\bibinfo  {journal} {Phys. Rev. Phys. Educ. Res.}\
  }\textbf {\bibinfo {volume} {17}},\ \bibinfo {pages} {020128} (\bibinfo
  {year} {2021})}\BibitemShut {NoStop}%
\bibitem [{\citenamefont {Dounas-Frazer}\ and\ \citenamefont
  {Lewandowski}(2017)}]{dounas2017electronics}%
  \BibitemOpen
  \bibfield  {author} {\bibinfo {author} {\bibfnamefont {D.~R.}\ \bibnamefont
  {Dounas-Frazer}}\ and\ \bibinfo {author} {\bibfnamefont {H.~J.}\ \bibnamefont
  {Lewandowski}},\ }\bibfield  {title} {\bibinfo {title} {Electronics lab
  instructors' approaches to troubleshooting instruction},\ }\href
  {https://doi.org/10.1103/PhysRevPhysEducRes.13.010102} {\bibfield  {journal}
  {\bibinfo  {journal} {Phys. Rev. Phys. Educ. Res.}\ }\textbf {\bibinfo
  {volume} {13}},\ \bibinfo {pages} {010102} (\bibinfo {year}
  {2017})}\BibitemShut {NoStop}%
\bibitem [{\citenamefont {Phillips}\ \emph {et~al.}(2021)\citenamefont
  {Phillips}, \citenamefont {Sundstrom}, \citenamefont {Wu},\ and\
  \citenamefont {Holmes}}]{phillips2021not}%
  \BibitemOpen
  \bibfield  {author} {\bibinfo {author} {\bibfnamefont {A.~M.}\ \bibnamefont
  {Phillips}}, \bibinfo {author} {\bibfnamefont {M.}~\bibnamefont {Sundstrom}},
  \bibinfo {author} {\bibfnamefont {D.~G.}\ \bibnamefont {Wu}},\ and\ \bibinfo
  {author} {\bibfnamefont {N.~G.}\ \bibnamefont {Holmes}},\ }\bibfield  {title}
  {\bibinfo {title} {Not engaging with problems in the lab: Students'
  navigation of conflicting data and models},\ }\href
  {https://doi.org/10.1103/PhysRevPhysEducRes.17.020112} {\bibfield  {journal}
  {\bibinfo  {journal} {Phys. Rev. Phys. Educ. Res.}\ }\textbf {\bibinfo
  {volume} {17}},\ \bibinfo {pages} {020112} (\bibinfo {year}
  {2021})}\BibitemShut {NoStop}%
\bibitem [{\citenamefont {Odden}\ \emph {et~al.}(2023)\citenamefont {Odden},
  \citenamefont {Silvia},\ and\ \citenamefont
  {Malthe-S{\o}renssen}}]{odden2023using}%
  \BibitemOpen
  \bibfield  {author} {\bibinfo {author} {\bibfnamefont {T.~O.~B.}\
  \bibnamefont {Odden}}, \bibinfo {author} {\bibfnamefont {D.~W.}\ \bibnamefont
  {Silvia}},\ and\ \bibinfo {author} {\bibfnamefont {A.}~\bibnamefont
  {Malthe-S{\o}renssen}},\ }\bibfield  {title} {\bibinfo {title} {Using
  computational essays to foster disciplinary epistemic agency in undergraduate
  science},\ }\href {https://doi.org/10.1002/tea.21821} {\bibfield  {journal}
  {\bibinfo  {journal} {Journal of Research in Science Teaching}\ }\textbf
  {\bibinfo {volume} {60}},\ \bibinfo {pages} {937} (\bibinfo {year}
  {2023})}\BibitemShut {NoStop}%
\bibitem [{\citenamefont {Phillips}\ \emph {et~al.}(2017)\citenamefont
  {Phillips}, \citenamefont {Watkins},\ and\ \citenamefont
  {Hammer}}]{phillips2017problematizing}%
  \BibitemOpen
  \bibfield  {author} {\bibinfo {author} {\bibfnamefont {A.~M.}\ \bibnamefont
  {Phillips}}, \bibinfo {author} {\bibfnamefont {J.}~\bibnamefont {Watkins}},\
  and\ \bibinfo {author} {\bibfnamefont {D.}~\bibnamefont {Hammer}},\
  }\bibfield  {title} {\bibinfo {title} {Problematizing as a scientific
  endeavor},\ }\href {https://doi.org/10.1103/PhysRevPhysEducRes.13.020107}
  {\bibfield  {journal} {\bibinfo  {journal} {Phys. Rev. Phys. Educ. Res.}\
  }\textbf {\bibinfo {volume} {13}},\ \bibinfo {pages} {020107} (\bibinfo
  {year} {2017})}\BibitemShut {NoStop}%
\end{thebibliography}%

\clearpage


\section*{Supplementary material}

# Affordances and Challenges of Incorporating a Remote, Cloud-accessible Quantum Experiment into Undergraduate Courses: Supplemental Material

In this supplement, we provide the questions analyzed for each of the data sources (Sec. I) as well as the details of how we classified the uncategorized courses in the survey data (Sec. II). For the instructor survey in Sec. I A, the possible responses (when not open-ended) are in italics in parentheses after the questions. Any time the option *Other* was available, the participants were provided an open-response box where they could briefly describe their choice.

## I. DATA SOURCES

### A. Instructor survey

Infleqtion's [Oqtant] is a remotely accessible Bose Einstein Condensate (BEC) experimental platform where anyone can create and manipulate quantum matter. Using the online interface, students have the opportunity to vary parameters related to (1) the final cooling stage during which the atoms condense into a BEC and (2) optical potentials the BEC can interact with. Students can input parameters and then view images of real experimental data using the parameters they entered. A sample screen shot of the data returned from a job is shown below, including real images of atoms, a fit to the data, and an extracted atom number. Further information can be found here or you can try out [Oqtant] yourself here (note that the system is not currently online, but you can still get a feel for it).

1. Please enter your name.

2. Please enter the name of your institution.

3. If you were provided with materials (e.g., activities for the students, an instructor's guide with explanations for how to use the activities, etc.), would you consider incorporating activities using [Oqtant] into a course you teach? (*Yes, Maybe, No*)

For the rest of the questions on this survey, please think about a specific course you teach in which you'd consider implementing [Oqtant]. If you teach multiple such courses and your responses to the survey questions would vary by course, feel free to fill out and submit this survey multiple times.

4. Please list the course number and name for the course you are responding about.

5. How would you classify this course? (*Beyond-first-year lab course, Quantum mechanics course, Other*)

    If *Beyond-first-year lab course:*

    (a) Approximately how many lab sessions during the course would you have to dedicate to an activity with [Oqtant]? (*1, 2, 3, 4, Other*)

    (b) How long is each of these sessions? (*1 hour, 2 hours, 3 hours, Other*)

    (c) How much time in total outside of the scheduled lab sessions would you want students to additionally spend working with [Oqtant]?

    If *Quantum mechanics course* or *Other*:

    (d) How much time during your course would you be willing to dedicate to students working with [Oqtant]?

6. There are many learning goals related to [Oqtant] for which we could design activities. Which of the following would you be most interested in incorporating into your course? Please select only the ones you think you would have time for. These options include both descriptions of possible learning goals and the way students would interact with [Oqtant] to accomplish those goals. Note that some of them involve students directly working with [Oqtant], whereas others would solely provide relevant background information. (Select all that apply.)

    - Students will be able to describe what is a BEC and explain how it differs from classical matter. This would involve conceptual activities mostly unrelated to [Oqtant].

- Students will be able to identify the different parts of the experimental apparatus used to create a BEC and explain the purpose of each of them. This would use [Oqtant] as an example, but would not involve direct student interaction with the experiment.

- Students will be able to explain conceptually how lasers can be used to cool atoms in different ways. This would use the various cooling stages in [Oqtant] as an example, but would focus on the general ideas.

- Students will be able to optimize the evaporative cooling parameters to create a BEC with [Oqtant]. They will be able to explain how the parameters affect the number and final temperature of the atomic cloud, and what these parameters imply about the density and collision rate of the atoms during evaporation. This would involve students controlling [Oqtant] through the online GUI and performing many cycles of choosing parameters and looking at the resulting data.

- Students will be able to analyze data coming from absorption images of atoms. They will be able to clean the images (e.g., remove fringes), and fit the data to extract the atom temperature and condensate fraction. This will involve activities using the data obtained from [Oqtant].

- Students will be able to explain conceptually how to use light to trap or repel atoms. This would use the optical potentials included in [Oqtant] as an example, but would focus on the general ideas.

- Students will use [Oqtant] to demonstrate that BECs exhibit quantum behavior (e.g., interference or collective behavior). This would involve students controlling [Oqtant] through the online GUI and performing many cycles of choosing parameters and looking at the resulting data.

- Students will be able to articulate and discuss various applications of BECs (e.g., quantum sensing) and ways they are used in modern research. This would involve conceptual activities mostly unrelated to [Oqtant], but drawing on knowledge students learned while interacting with [Oqtant].

- Other

7. Rank the importance of each of the following reasons for incorporating [Oqtant] into your course. (*Very important*, *Somewhat important*, *Not important* for each)

- Help students learn concepts about quantum mechanics.
- Prepare students for the quantum workforce.
- Provide students the opportunity to work with research-level experiments.
- Generate student excitement or appreciation for quantum physics (or physics more broadly).
- Other

8. Which of the following other experiments do you have in your course (or in other courses at your institution) that could provide students a hands-on experience complementary to working with [Oqtant]? [*Magneto-optical trap (MOT)*, *Saturated absorption spectroscopy*, *Other*]

9. What, if anything, will your students already know about BECs before working with [Oqtant]?

10. Is there anything you would be concerned about in terms of incorporating activities based on [Oqtant] into your course?

11. What kind of support would you need to feel comfortable incorporating [Oqtant] into your course?

12. Please enter anything else you'd like to share with us.

## B. Instructor interview protocol

1. How long have you been an instructor?

2. What range of courses have you taught?

3. What is your current research area?

4. What was your level of knowledge about BECs prior to deciding to incorporate Oqtant into your course?

5. What was your level of knowledge about atomic physics experiments prior to deciding to incorporate Oqtant into your course?

6. What was your level of knowledge about Python prior to deciding to incorporate Oqtant into your course?

7. What course did you use Oqtant in?

8. What level are the students in your course?

9. Have you taught this course before

    (a) If so, how frequently?
    (b) If so, how often do you adjust the materials in this course?

10. Why did you choose to implement activities with Oqtant in your course?

11. What specific learning goals did you have for using Oqtant?

12. What prior background knowledge were you assuming students had before this course (related to Oqtant)?

13. How did you implement Oqtant in your course?

    (a) Was it synchronous or asynchronous?
    (b) How much time did students have to work on the activities?
    (c) Did you include the preparation activities? If so, how?
    (d) Did it line up with any other content in your course?
    (e) Did students work alone or together?
    (f) Did the students work with Oqtant outside of the structured activities?
    (g) How did you grade/evaluate the activities with Oqtant?
    (h) How much of the students' final grade did activities with Oqtant contribute to?

14. Did you coordinate with other instructors at your institution when using Oqtant in your course?

15. How frequently did you communicate with me and/or staff at Infleqtion throughout this process? What method of communication did you use?

16. Were these activities at an appropriate physics-content level for your students?

17. Were these activities at an appropriate level of Python for your students?

18. Do you think your students achieved the learning goals you had for them? Why do you think that?

19. How do you think your students perceived these activities?

20. What skills do you think your students gained from working with Oqtant? How do you know this?

21. What knowledge do you think your students gained from working with Oqtant? How do you know this?

22. Are there any other benefits you think your students gained from working with Oqtant?

23. How did you feel about the progress the students made on their projects? Was it what you expected?

24. Overall, how was the process of using Oqtant in your class?

25. Did the activities cover the material you had hoped that they would?

26. Did the preparation activities cover the material you had hoped they would to prepare students for the activities?

27. Did the activities last a reasonable amount of time? Would you have wanted them to be shorter or longer?

28. How did the division of materials (e.g., the break point between Activity 1 and Activity 2) work?

29. Did you feel that you were successful at helping your students work through these activities?

30. How was communication with me and Infleqtion throughout the process?

31. Is there any other form of support you would have liked?

32. Did you feel like the activities with Oqtant fit well into the rest of your course?

33. Did you enjoy incorporating Oqtant into your course? Why or why not?

34. What challenges did you encounter while implementing Oqtant in your course? Were any of these particularly frustrating?

35. Is there anything you would do to change the way you implemented Oqtant if you were to use it again?

36. Would you choose to implement Oqtant in this course in the future? Why or why not?

37. Would you choose to implement Oqtant in another course you teach in the future? Why or why not?

38. What would you tell future instructors considering implementing Oqtant in their courses?

39. What do you wish that you had known before implementing Oqtant in your course?

40. Is there anything else you would like to share with me about your experience working with Oqtant in your course?

41. Are there any questions you wish I had asked you in this interview?

## C. Student interview protocols

Questions after Activity 1

1. What is absorption imaging?

2. Explain how you can obtain properties of atoms from absorption images.

3. What parts of this activity did you enjoy?

4. What parts did you not enjoy?

5. Did you feel like you were working with a real experiment even though you could only access it remotely?

6. How did the fact that you could not see the physical apparatus affect your experience?

7. Is there anything else you would like to share with me about your experience working with this activity?

Questions after Activity 2

1. How can you create a BEC experimentally?

2. How can you experimentally verify that atoms are in a BEC?

3. Explain what RF evaporation is and how it can be used to cool atoms.

4. Describe what the "time of flight" is in an atomic physics experiment and how it affects the shape of the atoms you see in an image.

5. Describe an effect you could see with a BEC that you would not see with a thermal cloud of atoms.

6. What is quantum about a BEC?

7. What parts of this activity did you enjoy?

8. What parts did you not enjoy?

9. How did the fact that you could not see the physical apparatus affect your experience?

10. What benefits and drawbacks to working with a remote experiment did you notice today?

11. How does this experiment compare with other experiments you have interacted with in lab courses and/or research experiences in the past?

12. Is there anything else you would like to share with me about your experience working with this activity?

## D. Jupyter notebook reflection questions

Activity 1

1. How does the kind of data analysis you performed in this activity compare with other kinds of data analysis you have done in prior labs? What are the similarities and differences?

2. What benefits and drawbacks do you notice so far to working with a remote experiment?

Activity 2

1. What, if any, parts of this activity emphasized the quantum nature of matter?

2. How similar or different do you think Oqtant is from other apparatus in research labs?

3. Did this activity feel like you were working with a real experiment even though you could only access it remotely? Explain why or why not.

## II. DATA ANALYSIS

For the closed-response parts of the instructor survey, we performed only one data cleaning step, which ensured the classification of the course types was consistent. Some of the instructors who chose *Other* as the type of course (as opposed to a BFY lab course or a quantum mechanics course), provided additional details about their courses, so we chose to classify some as BFY lab courses because they were a subset of that. The courses remaining in the *Other* category contain undergraduate courses in modern physics, thermodynamics and statistical mechanics, and a senior project in addition to one graduate theory course. Additionally, all responses but one were for a single course (or a sequence of similar courses); however there was one response that included multiple courses, including both BFY lab and quantum mechanics courses. Because we separated responses by type of course, we duplicated that response, assigning one as a BFY lab course and the other as a quantum mechanics course. Each data point therefore corresponds to one type of course taught by an individual instructor, leading to a dataset of instructor views about 31 total courses.



\end{document}